\documentclass[twocolumn,numberedappendix,iop]{openjournal}
\usepackage{lineno}
\usepackage{graphicx,amsmath,amssymb,amstext}
\usepackage{amsbsy,amsfonts,amsthm,color}
\usepackage[colorlinks,linkcolor=blue,citecolor=blue,urlcolor=blue]{hyperref}
\usepackage[utf8]{inputenc}
\usepackage{float}
\usepackage{xcolor}
\usepackage[T1]{fontenc}
\usepackage[title]{appendix}
\usepackage{savesym}
\savesymbol{tablenum}
\restoresymbol{SIX}{tablenum}
\shorttitle{DESI DR1 PV: correlation function results}
\shortauthors{Turner et al.}

\begin{document}
\title{The DESI DR1 Peculiar Velocity Survey: growth rate measurements from galaxy and momentum correlation functions \vspace{-4em}}

%---------------AUTHOR LIST------------------
\author{
R. J.~Turner,$^{1,*}$
C.~Blake,$^{1}$
F.~Qin,$^{2}$
J.~Aguilar,$^{3}$
S.~Ahlen,$^{4}$
A.~J.~Amsellem,$^{5}$
J.~Bautista,$^{2}$
S.~BenZvi,$^{6}$
D.~Bianchi,$^{7,8}$
D.~Brooks,$^{9}$
A.~Carr,$^{10}$
E.~Chaussidon,$^{3}$
T.~Claybaugh,$^{3}$
A.~Cuceu,$^{3}$
A.~de la Macorra,$^{11}$
P.~Doel,$^{9}$
K.~Douglass,$^{6}$
S.~Ferraro,$^{3,12}$
A.~Font-Ribera,$^{13}$
J.~E.~Forero-Romero,$^{14,15}$
E.~Gaztañaga,$^{16,17,18}$
S.~Gontcho A Gontcho,$^{3,19}$
G.~Gutierrez,$^{20}$
J.~Guy,$^{3}$
H.~K.~Herrera-Alcantar,$^{21,22}$
K.~Honscheid,$^{23,24,25}$
C.~Howlett,$^{26}$
D.~Huterer,$^{27,28}$
M.~Ishak,$^{29}$
R.~Joyce,$^{30}$
R.~Kehoe,$^{31}$
A.~G.~Kim,$^{3}$
D.~Kirkby,$^{32}$
A.~Kremin,$^{3}$
O.~Lahav,$^{9}$
Y.~Lai,$^{26,33}$
C.~Lamman,$^{25}$
M.~Landriau,$^{3}$
L.~Le~Guillou,$^{34}$
A.~Leauthaud,$^{35,36}$
M.~E.~Levi,$^{3}$
M.~Manera,$^{37,13}$
A.~Meisner,$^{30}$
R.~Miquel,$^{38,13}$
J.~Moustakas,$^{39}$
A.~Muñoz-Gutiérrez,$^{11}$
S.~Nadathur,$^{17}$
N.~Palanque-Delabrouille,$^{22,3}$
W.~J.~Percival,$^{40,41,42}$
C.~Poppett,$^{3,43,12}$
F.~Prada,$^{44}$
I.~P\'erez-R\`afols,$^{45}$
C.~Ross,$^{26}$
G.~Rossi,$^{46}$
K.~Said,$^{26}$
E.~Sanchez,$^{47}$
D.~Schlegel,$^{3}$
M.~Schubnell,$^{27,28}$
J.~Silber,$^{3}$
D.~Sprayberry,$^{30}$
G.~Tarl\'{e},$^{28}$
B.~A.~Weaver,$^{30}$
P.~Zarrouk,$^{34}$
and H.~Zou$^{48}$ \\
{\it (Affiliations can be found after the references)}
}
\thanks{$^*$E-mail: ryan.turner@anu.edu.au}

%--------------------------------------------

%\handlingeditor{Excellent E Editor}

%\doi{10.1017/pasa.2020.32}

%\received {dd Mmm YYYY}
%\revised  {dd Mmm YYYY}
%\accepted {dd Mmm YYYY}
%\published{22 September 202X}

%\keywords{Will be entered during the publication process} %% First letter not capped

\begin{abstract}
Joint analysis of the local peculiar velocity and galaxy density fields offers a promising route to testing cosmological models of gravity.  We present a measurement of the normalised growth rate of structure, $f\sigma_8$, from the two-point correlations of velocity and density tracers from the DESI DR1 Peculiar Velocity and Bright Galaxy Surveys, the largest catalogues of their kind assembled to date.  We fit the two-point correlation measurements with non-linear correlation function models, constructed from density and momentum power spectra generated using 1-loop Eulerian perturbation theory, and validate our methodology using representative mock catalogues.  We find $f\sigma_8 = 0.391^{+0.080}_{-0.081}$, consistent to within $1\sigma$ with accompanying analyses of the same datasets using power spectrum and maximum-likelihood fields methods.  Combining these growth rate results from different methods including appropriate correlations, we find a consensus determination $f\sigma_8(z = 0.07) = 0.4497 \pm 0.0548$, consistent with predictions from \textit{Planck}$+\Lambda$CDM cosmology.  Jointly fitting to this consensus low-redshift growth rate and the DESI DR1 full-shape clustering dataset, we measure gravitational growth index $\gamma_{\rm L} = 0.580^{+0.110}_{-0.110}$, consistent with the prediction of general relativity.
\\[1em]
\end{abstract}

\section{Introduction}
\label{sec:int}

The field of cosmology is currently grappling with several fundamental `tensions' that have the potential to change our understanding of the universe \citep[e.g.,][]{2016ApJ...826...56R, 2017MNRAS.467.3024L, 2018MNRAS.474.4894J, 2021APh...13102605D, 2022JHEAp..34...49A}.  It is often suggested that these problems will be resolved once newer, larger datasets become available, yet these issues seem to persist.  Indeed, supernova measurements from the Dark Energy Survey \citep[DES;][]{DES2024} and recent Baryon Acoustic Oscillations (BAO) measurements from the Dark Energy Spectroscopic Instrument \citep[DESI;][]{DESICollab2016, DESICollab2022} survey appear to favour a time-varying model of dark energy over the commonly-accepted cosmological constant \citep{DESIBAO2024, DESIBAO2025, Ishak2025}, posing further questions.

The local universe is a powerful laboratory for testing this dark energy component, which dominates the local cosmic dynamics.  Whilst the local expansion is chiefly sensitive to the Hubble constant, and the local volume is insufficient for accurate galaxy clustering measurements, the local motions of galaxies nonetheless allow accurate tests of gravitational physics.  These so-called `peculiar velocities' are induced in galaxies by the gravitational influence of nearby large-scale structure, causing departures from their expected recessional velocities. These effects modify the cosmological redshift of a galaxy as:
\begin{equation}
\label{eq:redshift}
    (1 + z) = (1 + \bar{z})\left(1 + \frac{v_p}{c}\right) \, ,
\end{equation}
where $z$ is the measured redshift, $\bar{z}$ is the cosmological redshift, and $v_p$ is the peculiar velocity \citep[for a discussion of fundamentals, see][]{Davis2014}. Through comparing the spectroscopically-obtained redshift of a galaxy, $z$, to the inferred Hubble-flow redshift for that galaxy obtained from a redshift-independent distance indicator, $\bar{z}$, we are able to infer the peculiar velocity of the galaxy, $v_p$.  Typically, the Fundamental Plane Relation \citep[FP;][]{Djorgovski1987, Dressler1987} and the Tully-Fisher Relation \citep[TF;][]{Tully1977} are used in large galaxy samples to measure redshift-independent distances.  For a review of peculiar velocity cosmology, see \cite{Turner2025}.

As part of its 8-year observing program, DESI is assembling the best currently-existing spectroscopic datasets of the local Universe, including a dedicated survey of peculiar velocities.  This survey is expected to obtain $\sim 180{,}000$ peculiar velocities by the end of the observing program \citep{Saulder2023}, roughly five times larger than the previous largest such homogeneously-selected catalogue, the Sloan Digital Sky Survey Peculiar Velocity Survey \citep[SDSS PV;][]{Howlett2022}.  This paper is one of a set describing the analysis of the first year of data from the DESI Peculiar Velocity survey, accompanying Data Release 1 \citep[DR1,][]{2025arXiv250314745D}. In this work we will describe the measurement of the growth rate of structure from galaxy and momentum (density-weighted velocity) correlation function statistics.

The growth rate of structure, $f$, describes the logarithmic growth of density perturbations $D$, with respect to the cosmic scale factor $a$, and can also be described by an empirical fitting formula,
\begin{equation}
    f \equiv \frac{d \ln D(a)}{d\ln a} \simeq \Omega_m(a)^{\gamma} \, ,
\label{eq:gamma}
\end{equation}
where $\Omega_m(a)$ is the time-dependent matter density parameter, and $\gamma$ is the growth index (as defined by \citealt{Linder2005}, but we also note the earlier work of \citealt{Wang1998}), where $\gamma_{GR} = 0.55$ for a General Relativity theory of gravity.  \cite{Linder2005} also shows that this empirical relation is precise even when extended to cosmologies with dynamical dark energy.  Cosmological models that share an expansion history, but have differing gravitational theories, will find different values for $\gamma$.  Since the amplitude of peculiar velocities is also driven by the amplitude of density fluctuations in the Universe, momentum correlations probe (at leading order) the combined parameter $f\sigma_8$, the normalised growth rate of large-scale structure.  The parameter $\sigma_8$ describes the root mean square amplitude of linear mass fluctuations on scales of $R_8 = 8\,h^{-1}$Mpc, measured at the present time $z = 0$,
\begin{equation}
    \sigma_8^2 = \frac{1}{2\pi^2}\int_0^{\infty}dk\,k^2\,P(k)\,|W(kR_8)|^2 \, ,
\end{equation}
where $P(k)$ is the linear matter power spectrum measured at the present time and $W(kR_8)$ is the Fourier transform of the spherical top hat window function with radius $R_8$.

%To directly measure peculiar velocities we require redshift-independent distance measurements. Measuring $\bar{z}$ in this way, in conjunction with accurate spectroscopic measurements of the observed redshift $z$, enables us to infer $v_p$ directly from Equation \ref{eq:redshift}.  The DESI Peculiar Velocity survey uses the Fundamental Plane Relation \citep[FP;][]{Djorgovski1987, Dressler1987} and the Tully-Fisher Relation \citep[TF;][]{Tully1977} to measure redshift-independent distances.  For a review of peculiar velocity cosmology, see \cite{Turner2025}.

Peculiar velocities also distort the redshift-space positions of galaxies, and hence induce anisotropies in the galaxy clustering signal as a function of the angle to the line-of-sight \citep{Kaiser1984, Kaiser1987}, a phenomenon known as redshift-space distortions (RSDs).  Growth rate information can be extracted from RSDs by measuring the clustering signal in redshift space.  Directly measured peculiar velocities and statistically inferred peculiar velocities are complementary probes. The Fourier modes of the momentum and density fields are related by $\tilde{v}(k) \propto \tilde{\delta}(k)/k$, causing an increase in signal-to-noise for velocities at larger scales (smaller $k$), such that peculiar velocities are organised in `bulk flows'. The momentum power spectrum hence traces the underlying matter power spectrum on large scales, where we expect to see evidence for modified gravity models before they are screened, whilst the signal-to-noise of RSD is dominated by smaller-scale modes \citep{Koda2014}.
  
In recent years, significant effort has been invested in producing methods for extracting cosmological information from peculiar velocities. Under the assumption that the momentum and galaxy density fields are Gaussian, we can use the two-point summary statistics between velocity and density tracers to measure the growth rate; this has been done with the Fourier-space power spectra \citep{Park2000, Howlett2017b, Qin2019, Qin2025} and the configuration-space correlation functions \citep{Nusser2017, Wang2018, Dupuy2019, Courtois2023, Turner2023, Lyall2024}. Maximum likelihood methods that analytically model the covariance between the velocity and galaxy density fields, assuming both fields are correlated samples drawn from multivariate Gaussians, have also been used to constrain the growth rate by \cite{Johnson2014, Huterer2017, Adams2020}, and \cite{Lai2023}. Reconstruction techniques, where the velocity field is inferred from the measured galaxy density field and then directly compared to discrete velocity measurements point-by-point, are another viable method of constraining the growth rate \citep{Davis2011, Turnbull2012, Carrick2015, Said2020, Lilow2021, Qin2023, Boubel2024, Stiskalek2025}. We refer readers to \cite{Strauss1995} for a historical review of the science achievable with the local galaxy density and peculiar velocity fields.

A series of companion papers present the DESI DR1 PV datasets, and complementary cosmological analyses.  Measurements of $f\sigma_8$ from the power spectra are described by \cite{FeiPV}, and results from the maximum-likelihood method are described by \cite{YanPV}. The datasets themselves are described by \cite{CaitlinPV} who cover the DESI DR1 Fundamental Plane catalogue, and \cite{KellyPV} who present the DESI DR1 Tully-Fisher catalogue.  The zero-point calibration for both catalogues, as well as constraints on the Hubble constant $H_0$, are described by \cite{AnthonyPV}. The galaxy and velocity mocks produced for this analysis are described by \cite{JulianPV}.  We will refer readers to these papers as appropriate.

Our paper is structured as follows: in Section \ref{sec:DESI} we describe the DESI DR1 and mock datasets we utilise in this work. In Section \ref{sec:fitting} we describe the construction of the models and estimators used in our analysis, and the methodology we use to fit our models to the data.  We present our fitting results on the mocks and data, and discuss our findings, in Section \ref{sec:results}.  Finally, we conclude and propose ideas for future work in Section \ref{sec:conclusion}.

\section{The Dark Energy Spectroscopic Instrument}
\label{sec:DESI}

The Dark Energy Spectroscopic Instrument is a multi-fibre spectrograph installed on the 4m Mayall Telescope at the Kitt Peak National Observatory, Arizona, USA. The DESI survey is a spectroscopic survey designed to map the large-scale structure of the universe over a wide redshift range and substantial cosmological volume, with the goal of measuring the expansion history of the universe as well as the growth rate of large-scale structure \citep{2013arXiv1308.0847L, DESICollab2016, 2016arXiv161100037D}. The DESI Survey is planned to span eight years, during which it will observe more than 60 million galaxies and quasars over a footprint of $17{,}000$ square degrees.  The survey is internally divided into four main target classes: the Bright Galaxy Survey (BGS), the Luminous Red Galaxy Survey (LRG), the Emission Line Galaxy Survey (ELG), and the Quasar Survey (QSO), where targets are all selected from the DESI Legacy optical imaging surveys \citep{2019AJ....157..168D}.

The DESI focal plane covers 8 square degrees and contains 5000 optical fibers \citep{FiberSystem.Poppett.2024}, which can be individually placed onto targets in each observational field using a robotic positioner \citep{2023AJ....165....9S} and an optical corrector \citep{Corrector.Miller.2023}. In each exposure, the observed targets are measured across the wavelength range $3600 - 9800$\AA{} and the spectral data are processed by Redrock\footnote{\url{https://github.com/desihub/redrock}}, the DESI spectroscopic pipeline \citep{2023AJ....165..144G}.  The overall observing strategy is summarised by \cite{2023AJ....166..259S}.  DESI has issued an Early Data Release \citep{2024AJ....168...58D}, which has been used for scientific validation of the program \citep{DESI-SV}. For this work we use a galaxy sample drawn from the publicly available DESI Data Release 1 \citep{2025arXiv250314745D}, specifically the DESI DR1 large-scale structure catalogues\footnote{\url{https://data.desi.lbl.gov/doc/releases/dr1/}}\citep{desi-dr1-sample}. The velocity sample is taken from the first year of observations of the DESI Peculiar Velocity (DESI PV) survey, as described by \cite{CaitlinPV} and \cite{KellyPV}.  We briefly summarise these datasets below.

\subsection{The Bright Galaxy Survey dataset}

In our study, we wish to jointly analyse the clustering statistics of the momentum and galaxy density fields.  The momentum field is only well measured at redshifts $z \lesssim 0.1$, as this is the range over which distance errors are not yet dominant in estimates of galaxy peculiar velocities, the intrinsic scatter in the galaxy scaling relations are not too large, and the fractional contribution of peculiar velocities to the measured redshift of sources is still significant. Thus, we use redshift data from the DESI Bright Galaxy Survey (BGS), the lowest-redshift DESI sample, to measure galaxy clustering statistics. The BGS sample is a magnitude-limited sample of galaxies with $r$-band magnitudes $14 < r < 19.5$ out to redshift $z < 0.5$, and BGS targets are assigned high priority during DESI bright-time observations. The full selection criteria and validation of the BGS sample are described by \cite{2023AJ....165..253H}.

The final BGS DR1 sample has a density of 854 deg$^{-2}$ containing reliable redshift measurements for over 5.5 million galaxies, and consists of a magnitude-limited Bright sample with $r < 19.5$ and a colour-selected Faint component with $19.5 < r < 20.175$. The Faint sample suffers from complications with regard to incompleteness and systematics \citep{2023AJ....165..253H}, and so we solely focus on the Bright sample in this work.  To overlap with our peculiar velocity datasets we restrict our analysis to the redshift range $z < 0.1$.  To minimize redshift evolution across the sample, which might complicate our interpretation of clustering statistics, we further apply a luminosity cut based on the absolute magnitude in the $r$-band $M_r < -17.7$, which creates a sample with a roughly constant number density across $z < 0.1$ containing $415{,}523$ galaxies.

\cite{2025JCAP...01..125R} describes how large-scale structure catalogues appropriate for cosmological analyses have been constructed from the redshift and target catalogues of each DESI tracer. The corresponding sample selection function is defined, and weights are designed to compensate for systematic density variations associated with target selection systematics, spectroscopic incompleteness, and fibre collisions. Corresponding random catalogues are generated from these selection functions, which are used in estimators of the galaxy two-point correlation function.  We also use these random catalogues in our analysis, sub-sampled in redshift to match the additional luminosity cut we have applied to the data.

\subsection{DESI PV survey}
\label{ssec:pvcat}

A typical DESI pointing will leave some fibres with no available target, and so secondary targeting programs have been created to make use of these unused fibres. The DESI Peculiar Velocity survey is one such secondary program, designed to measure galaxy peculiar velocities in the local universe. It is the first survey of its kind to measure peculiar velocities using both the Tully-Fisher (TF) relation for late-type spiral galaxies and the Fundamental Plane (FP) relation for early-type elliptical galaxies. This versatility means that, once completed, the DESI PV survey will be the largest and most homogeneous catalogue of peculiar velocities ever collected.

The DESI PV survey target selection and characteristics are fully described by \cite{Saulder2023}.  Over the 8-year span of the DESI survey, the PV program will measure more than 180,000 velocities across the FP and TF samples, with a forecast growth rate measurement with $4\%$ statistical error.  \cite{FP-EDR} and \cite{TF-EDR} describe the calibration of the FP and TF samples using spectroscopic data from the DESI Survey Validation sample \citep[SV;][]{DESI-SV}, while \cite{CaitlinPV} and \cite{KellyPV} describe the respective DR1 FP and TF samples.

\subsubsection{FP catalogue}

The Fundamental Plane relation is a method of determining redshift-independent distance measurements for early-type elliptical galaxies. The FP relation is an empirical galaxy scaling relation that links the kinematic properties of elliptical bulge-dominated galaxies to their structural properties. The relation is given by,
\begin{equation}
    \log R_e = a\log\sigma_o + b\log I_e + c  \, ,
\end{equation}
where $R_e$ is the effective radius of the galaxy, $I_e$ is the surface brightness interior to the effective radius, and $\sigma_o$ is the central velocity dispersion. The coefficients $a$ and $b$ are the slopes of the plane, while $c$ describes the offset of the plane.

After selection and quality cuts are applied, the FP DR1 catalogue contains $73{,}822$ elliptical galaxies in the redshift range $0.01 < z < 0.1$. There are a small fraction of outliers in the log-distance ratio values that cause the distribution of log-distance ratios to be skewed negative. To address this, we remove any galaxies from the catalogue whose log-distance ratio are more than $4\sigma$ from the median.  This removes 136 objects from the FP catalogue, leaving $73{,}686$ elliptical galaxies for our analysis.

\subsubsection{TF catalogue}

The Tully-Fisher relation is a method for determining redshift-independent distances of late-type spiral galaxies. The TF relation links the rotational velocity of spiral galaxies to their absolute magnitude, as a proxy for luminosity. The apparent magnitude of a galaxy is a distance-dependent observable, and so we can contrast this with the predicted absolute magnitude from the TF relation to determine the distance to the galaxy. The TF relation is given by,
\begin{equation}
    M = a\log V + b  \, ,
\end{equation}
where $M$ is the absolute magnitude of the galaxy and $V$ is the rotational velocity. The coefficients $a$ and $b$ describe the slope and zero-point of the relation, respectively.

The TF relation is typically constructed using the neutral hydrogen (HI) 21-cm line at radio wavelengths, as HI is a good tracer of the rotational velocity of galaxies.  As DESI is an optical survey, we instead rely on H$\alpha$ emission to trace rotational velocity. DESI fibres are placed on the centres and semi-major axes of extended spiral galaxies, from which the rotational velocity of the galaxies can be inferred. The DR1 Tully-Fisher catalogue contains $6{,}806$ spiral galaxies over the same redshift range as the Fundamental Plane catalogue. 

\begin{figure}
    \centering
    \includegraphics[width=\linewidth]{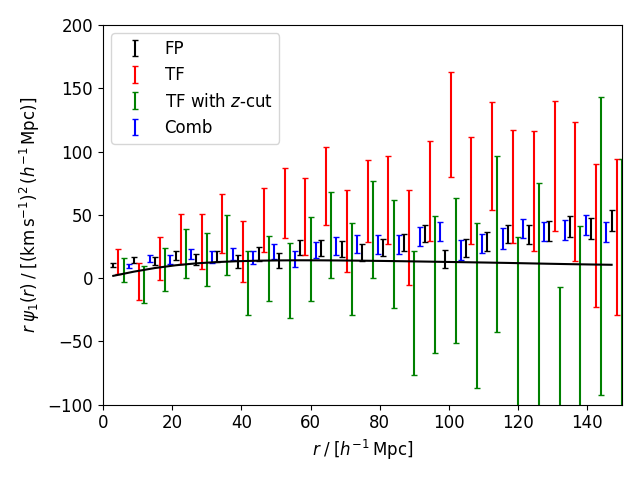}
    \caption{The $\psi_1$ correlation function for different velocity datasets tested in our analysis: the FP sample (black points), the TF sample (red points), the TF sample with a $z < 0.05$ cut (green points), and the combined sample (blue points).  The solid line is the fiducial model.  The original TF sample shows a high amplitude of correlation, leading us to apply a redshift cut $z < 0.05$ for the cosmological analysis.}
    \label{fig:psi1corr_comparison}
\end{figure}

During the correlation measurement process, we identified an excess correlation between the measured velocities within the TF peculiar velocity dataset at $z > 0.05$.  We illustrate this effect in Figure \ref{fig:psi1corr_comparison}, where we compare measurements of the $\psi_1$ momentum correlation function (introduced in Section \ref{ssec:estim}) from the full TF sample, the TF sample cut at $z < 0.05$, the FP dataset, and the combination of the FP and $z$-cut TF datasets.  The reasons for this excess correlation (which is also seen in the other analysis methods) are yet to be understood, but in order to be conservative we only considered the $z < 0.05$ component of the TF sample in our analysis, reducing the sample size to $2{,}934$ galaxies (although we note that the TF sample does not significantly contribute to the growth rate determination at this stage of the survey, so this cut has no effect on the conclusions of our study).  Applying the $4\sigma$ log-distance ratio cut to this subsample removes an additional 5 galaxies, leaving $2{,}929$ spiral galaxies in our sample. 

Our final PV sample hence contains $73{,}686$ FP velocities and and $2{,}929$ TF velocities, for a total DR1 sample size of $76{,}615$. We choose the effective redshift of the sample to be the mean redshift of all galaxies across the clustering and velocity samples, $z_{\rm eff} = 0.07$.

\subsection{Mocks}
\label{sec:mocks}

We create simulated clustering and peculiar velocity catalogues closely matching the BGS and PV datasets used in this analysis. These mocks enable us to perform robust tests of our methodology before applying it to the DR1 data, and are fully described by \cite{JulianPV}.  The mocks are produced from 25 realisations of the {\sc AbacusSummit} simulation \citep{Garrison2021, Maksimova2021}, where each realisation is subsampled to produce 27 independent mock datasets.  Hence we use 675 clustering and velocity mocks, which we sub-sample to match the redshift distribution and angular selection of the real datasets, and apply peculiar velocity noise matching the FP and TF samples.  The {\sc AbacusSummit} simulations are produced with a $\Lambda$CDM cosmology with matter density $\Omega_m = 0.3153$, baryon density $\Omega_b = 0.0493$, Hubble parameter $h = 0.6736$, normalisation $\sigma_8 = 0.8114$, and spectral index $n_s = 0.9649$ \citep[the cosmology favoured by][]{PlanckCollab2018}.  Our mocks are based on snapshots generated at a redshift of $z = 0.2$, with a fiducial value of the normalised growth rate $f\sigma_8 = 0.462$. We note that this is different to the effective redshift of the DR1 dataset, but this is the lowest-redshift snapshot that was fit with a halo occupaton distribution (HOD) model. In future work we would like to re-fit the HOD parameters to the z = 0.1 {\sc AbacusSummit} snapshot, which is the lowest redshift snapshot available to us.

\section{Fitting procedure}
\label{sec:fitting}

\subsection{Models}
\label{ssec:models}

In this section, we describe the models we adopt for fitting the galaxy and momentum auto- and cross-correlations.  We utilise the 1-loop Eulerian standard perturbation theory approach of \cite{FeiPV} to generate non-linear models of the density and momentum power spectra, which we transform to non-linear correlation function models.  We will not describe in detail how these power spectra models are produced, and instead direct interested readers to the accompanying paper for a thorough description of the modelling.  Here, we describe the conversion of the power spectra to the correlation function models used in our analysis.  We utilise five correlation functions in total: two momentum auto-correlations, the monopole and quadrupole of galaxy clustering, and the cross-correlation dipole.

\subsubsection{Momentum auto-correlation model}

We start by describing our models for the momentum auto-correlation functions, $\psi_1$ and $\psi_2$ \citep{Gorski1989}.  These take the form,
\begin{equation}
    \label{eq:psi1}
    \psi_1(r) = A(r) \, \Psi_{\parallel}(r) + [1-A(r)] \, \Psi_{\perp}(r) \, ,
\end{equation}
\begin{equation}
    \label{eq:psi2}
    \psi_2(r) = B(r) \, \Psi_{\parallel}(r) + [1-B(r)] \, \Psi_{\perp}(r) \, ,
\end{equation}
where $\Psi_{\perp}(r)$ and $\Psi_{\parallel}(r)$ describe the transverse and radial correlations of the three-dimensional peculiar velocity field, and the terms $A(r)$ and $B(r)$ represent the geometry of the survey.  It has been shown that the radial velocity correlation estimators $\psi_1$ and $\psi_2$ contain all of the information within the 3D velocity tensor \citep{Gorski1988, Wang2018, BlakeCov}, under the assumption of an irrotational velocity field.

The functions $\Psi_{\perp}(r)$ and $\Psi_{\parallel}(r)$ are given by,
\begin{equation}
    \label{eq:psiperp}
    \Psi_{\perp}(r) = \frac{1}{2\pi^2} \int dk \, k^2 \, P_v(k) \,\left[j_0(kr) - 2\frac{j_1(kr)}{kr}\right] \, ,
\end{equation}
\begin{equation}
    \label{eq:psipara}
    \Psi_{\parallel}(r) = \frac{1}{2\pi^2} \int dk \, k^2 \, P_v(k) \, \frac{j_1(kr)}{kr} \, ,
\end{equation}
where $j_n(x)$ are the spherical Bessel functions, and we set $P_v(k) = 3 P^0_{pp}(k)$, where $P^0_{pp}(k)$ is the monopole of the line-of-sight momentum power spectrum produced by the code of \cite{FeiPV}.  The factor of 3 can be readily understood if we consider the leading-order model for the angle-dependent line-of-sight momentum power spectrum,
\begin{equation}
    \label{eq:momPS}
    %P_{pp}(k,\mu) = \left(\frac{aHf\mu}{k}\right)^2 \, D_u^2(k) \, P_L(k) \, ,
    P_{pp}(k,\mu) = \left(\frac{aHf\mu}{k}\right)^2 \, P_L(k) \, ,
\end{equation}
where $P_L(k)$ is the linear matter power spectrum and $\mu$ is the cosine of the angle between the vector $\vec{r}$, describing the separation of two galaxies at positions $\vec{s}_1$ and $\vec{s}_2$, and the line of sight vector $\vec{d}$. We choose $\vec{d}$ such that $\vec{d}\cdot\vec{s}_1$ = $\vec{d}\cdot\vec{s}_2$. This is also illustrated in Figure \ref{fig:geom}.
%and $D_u(k)$ is a non-linear velocity damping term given by
%\begin{equation}
%    D_u(k) = \frac{\sin(k\sigma_{v,2})}{k\sigma_{v,2}} \, ,
%\end{equation}
%where $\sigma_{v,2}$ is the velocity dispersion attributed to the velocity field.  
The monopole of Equation \ref{eq:momPS} is obtained by integrating over $\mu$, ranging from 0 to 1,
\begin{equation}
    %P^0_{pp}(k) = \frac{1}{3} \left( \frac{a H f}{k} \right)^2 D_u^2(k) P_L(k) = \frac{1}{3} P_v(k) \, ,
    P^0_{pp}(k) = \frac{1}{3} \left( \frac{a H f}{k} \right)^2 P_L(k) = \frac{1}{3} P_v(k) \, ,
\end{equation}
where $P_v(k)$ is the velocity power spectrum that appears in Equation \ref{eq:psiperp} and Equation \ref{eq:psipara}.

To convert from power spectrum multipoles to correlation function models, we utilise the python implementation of the {\sc FFTLog} algorithm within the package {\sc cosmoprimo}\footnote{https://github.com/cosmodesi/cosmoprimo}.
%we perform the Hankel transforms represented by Equation \ref{eq:psiperp} and Equation \ref{eq:psipara} using the {\sc{hankl}} software \citep{2021arXiv210606331K}, a python implementation of FFTLog.  
To perform these transforms robustly, we extend the power spectrum models from $k_{\mathrm{max}} \sim 0.4$ h Mpc$^{-1}$ to higher values of $k$ using power-law extrapolation.  We also apply an exponential damping to the power spectra of the form $\exp(-k^2 a^2)$, setting $a = 1\,h^{-1}$ Mpc, which does not affect any of the parameter fits but mitigates high-frequency oscillations in the outputs.  Finally, the geometric factors $A(r)$ and $B(r)$ that appear in Equation \ref{eq:psi1} and Equation \ref{eq:psi2} are given by sums over the data sample in each separation bin,
\begin{equation}
    \label{eq:amodel}
    A(r) = \frac{\sum_{a<b}^{|r_{ab}|\in \mathrm{bin}} W_{\eta,a}\,W_{\eta,b}\,\cos\theta_{ab}\,\cos\theta_a\,\cos\theta_b}{\sum_{a<b}^{|r_{ab}|\in \mathrm{bin}} W_{\eta,a}\,W_{\eta,b}\,\cos^2\theta_{ab}} \, ,
\end{equation}
\begin{equation}
    \label{eq:bmodel}
    B(r) = \frac{\sum_{a<b}^{|r_{ab}|\in \mathrm{bin}} W_{\eta,a}\,W_{\eta,b}\,\cos^2\theta_a\,\cos^2\theta_b}{\sum_{a<b}^{|r_{ab}|\in \mathrm{bin}} W_{\eta,a}\,W_{\eta,b}\,\cos\theta_a\,\cos\theta_b\,\cos\theta_{ab}}  \, ,
\end{equation}
where $W_{\eta,a}$ and $W_{\eta,b}$ denote the per-object weights (introduced in Section \ref{subsec:weighting}), and we are summing over all unique pairs ($a < b$) and adding to the bin total when the magnitude of the separation vector between galaxies $a$ and $b$ falls in that bin ($|r_{ab}| \in \mathrm{bin}$).

\subsubsection{Galaxy auto-correlation model}

We now describe the models for the galaxy auto-correlation functions.  We quantify the information in the galaxy density field using the monopole and quadrupole of the correlations as a function of the angle with respect to the line-of-sight, which are computed from the corresponding power spectrum multipoles as,
\begin{equation}
    \label{eq:ggmodel}
    \xi^{\ell\,=\,0,2}_{gg}(r) = \frac{i^{\ell}}{2\pi^2}\int dk \, k^2 \,  P^{\ell}_{gg}(k) \, j_{\ell}(kr) \, ,
\end{equation}
where $P^{\ell}_{gg}(k)$ are the multipoles of the galaxy power spectrum as produced by the code of \cite{FeiPV}.  The leading-order behaviour can be captured as,
\begin{equation}
    %P_{gg}(k,\mu) = (b_1 + f\mu^2)^2\,D^2_g\,P_L(k) \, ,
    P_{gg}(k,\mu) = (b_1 + f\mu^2)^2\,P_L(k) \, ,
\end{equation}
where $b_1$ is the linear galaxy bias which accounts for the fact that the distribution of galaxies, $\delta_g$, is a biased tracer of the underlying matter distribution, $\delta_g = b_1\delta$.
%and $D_g(k)$ is an additional non-linear damping term given by,
%\begin{equation}
%    D_g(k,\mu) = \left[1 + 0.5(k\mu\sigma_{v,1})^2\right]^{-1/2} \, ,
%\end{equation}
%where $\sigma_{v,1}$ is the non-linear velocity dispersion attributed to the density field.

\subsubsection{Galaxy-momentum cross-correlation model}

We also analyse the galaxy-momentum cross-correlation, whose leading-order behaviour can be expressed using the dipole,
\begin{equation}
    \label{eq:gumodel}
    \xi^{\ell\,=\,1}_{gu}(r) = \frac{1}{2\pi^2}\int dk\, k \, P^1_{gp}(k) \, j_1(kr) \, ,
\end{equation}
where $P^1_{gp}(k)$ is output by the modelling code of \cite{FeiPV}.  The octopole ($\ell = 3$) of the cross-correlation is, in theory, non-zero too, but we find the signal to be too low to warrant inclusion in this analysis.  The leading-order behaviour can be described by,
\begin{equation}
    %P_{gp}(k,\mu) = (b_1 + f\mu^2)\,D_g\,D_u\frac{iaHf\mu}{k}P_L(k)  \, .
    P_{gp}(k,\mu) = (b_1 + f\mu^2)\,\frac{iaHf\mu}{k}P_L(k)  \, .
\end{equation}
We again compute the transforms of Equation \ref{eq:ggmodel} and Equation \ref{eq:gumodel} using {\sc{CosmoPrimo}}.

\subsection{Estimators}
\label{ssec:estim}

We now describe the estimators of the five correlation functions we use in our analysis, following \cite{Turner2021, Turner2023}.

\subsubsection{Log-distance ratios}

As we will describe in Section \ref{ssec:minim}, our results are obtained from a likelihood analysis which requires our peculiar velocity data to be drawn from a multivariate Gaussian distribution with a mean of zero. Although this behaviour is found in linear theory, the presence of observational errors complicates matters. As shown by \citet{Springob2014}, the peculiar velocity errors obtained from the Fundamental Plane are lognormally-distributed. This is a direct result of the Gaussian error distribution for galaxy magnitude offsets from the Fundamental Plane, as described by Appendix A of \citet{Springob2014}. We can instead frame these data in terms of the logarithmic ratio between the comoving distance calculated from the observed redshift, $D(z_{\rm obs})$, and the true comoving distance estimated using the Hubble redshift, $D(z_{\rm H})$. The former we calculate from $z_{\rm obs}$ using a cosmological model, the latter is what we measure using the TF and FP relations. This is simply given by
\begin{equation}
    \eta \equiv \log_{10}D(z_{\rm obs}) - \log_{10}D(z_{\rm H}) = \log_{10}\left[\frac{D(z_{\rm obs})}{D(z_{\rm H})}\right] \, .
\end{equation}
The measured values of $\eta$ are Gaussian-distributed to a good approximation, so we recast our velocity measurements in terms of this parameter. The conversion between the logarithmic distance ratio $\eta$ and the peculiar velocity $v_p$ can be expressed as, 
\begin{equation}
    \label{eq:logd_to_vp}
    \eta = \alpha \, v_p \, ,
\end{equation}
where
\begin{equation}
\label{eq:alpha}
    \alpha(z_{\rm obs}) = \frac{1}{\ln(10)}\frac{1 + z_{\rm obs}}{D(z_{\rm obs})H(z_{\rm obs})} \, ,
\end{equation}
is a simple function of the observed redshift, following \citet{Johnson2014} and \citet{Adams2017} (see Appendix B of \citet{Adams2017} for a full derivation).

\begin{figure*}
    \centering
    \includegraphics[width=0.5\textwidth]{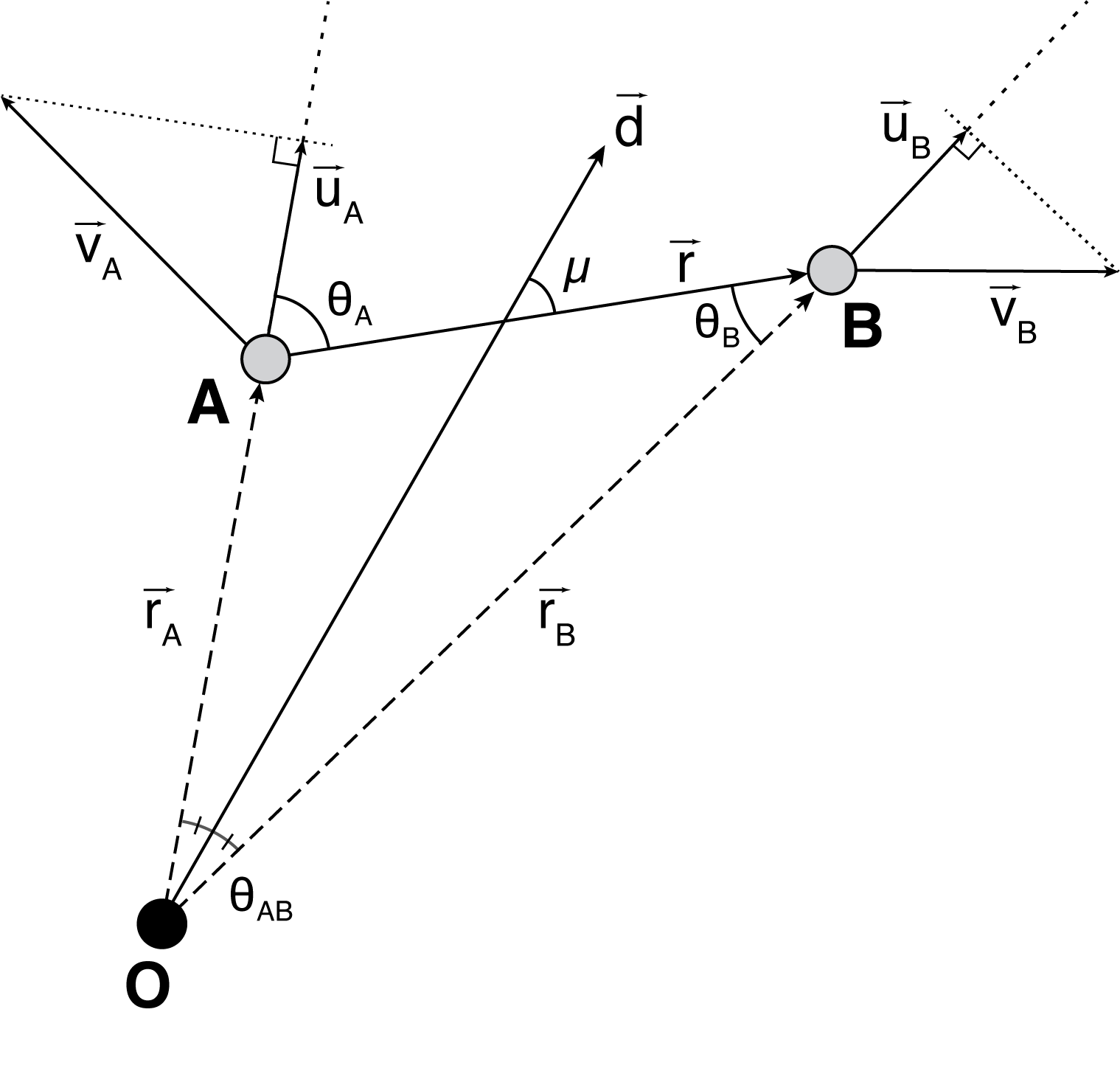}
    \caption{A visualisation of the geometry of two galaxies $A$ and $B$ in relation to an observer $O$. Vectors separating the observer from galaxies $A$ and $B$ are described by $\vec{r}_a$ and $\vec{r}_b$, respectively, while $\vec{r}$ is the vector separating $A$ and $B$. The three-dimensional velocity of the galaxies are represented by $\vec{v}_a$ and $\vec{v}_b$, while the measurable line of sight component of the velocity is given by $u_a$ and $u_b$. The cosines of the angles $\theta_a$, $\theta_b$, and $\theta_{ab}$ describe the relative locations of the observer and galaxy pair. The vector intersecting $\vec{r}$ that bisects the angle $\theta_{ab}$ is given by $\vec{d}$. The cosine of the angle between $\vec{r}$ and $\vec{d}$ is given by $\mu$. Adapted from a similar figure in \cite{Turner2025}.}
    \label{fig:geom}
\end{figure*}

\subsubsection{Optimal Weighting}
\label{subsec:weighting}
When measuring the galaxy and momentum correlation functions, we weight the objects in accordance with the optimal weighting scheme \citep{Feldman1994} that balances sample variance and noise.  For a galaxy $i$ in our galaxy density sample we use the weight,
\begin{equation}
    W_{g,\,i} = \frac{\kappa_i}{1 + n_i \, P_{gg}} \, ,
    \label{eq:ggfkp}
\end{equation}
where $\kappa_i$ is the completeness weight applied to account for variations in the selection function \citep[see Section 8.2 of][for a full description]{desi-dr1-sample}, $P_{gg} = 10^4 \, h^{-3}$ Mpc$^3$ is the characteristic amplitude of the galaxy power spectrum, and $n_i$ is the number density of the redshift sample at galaxy $i$. 

The optimal weight to be applied in the peculiar velocity sample (measured using the logarithmic distance ratio $\eta$), from \cite{Turner2023}, is
\begin{equation}
    \label{eq:newFKP}
    W_{\eta,\,i} = \frac{1}{\sigma_{\eta,i}^2/\alpha_i + \, \alpha_i
    n_i \, P_v} \, ,
\end{equation}
where $\alpha$ is the conversion factor defined in Equation \ref{eq:alpha}, $P_v = 10^{10} \, h^{-3}$ Mpc$^3$ km$^2$ s$^{-2}$ is the characteristic amplitude of the momentum power spectrum, and $n_i$ is the number density of the velocity sample at galaxy $i$.

\subsubsection{Momentum auto-correlation estimator}

The estimators of the momentum auto-correlation functions $\psi_1(r)$ and $\psi_2(r)$ are measured as a function of separation and take the form,
\begin{equation}
    \hat{\psi}_1(r) = \frac{\sum_{a<b\in{\rm DD}}^{|r_{ab}| \in \mathrm{bin}} W_{\eta,\,a}\,W_{\eta,\,b}\,\eta_a\,\eta_b\,\cos\theta_{ab}}{\sum_{a<b\in{\rm DD}}^{|r_{ab}| \in \mathrm{bin}} W_{\eta,\,a}\,W_{\eta,\,b}\,\cos^2\theta_{ab}} \, ,
\end{equation}
and,
\begin{equation}
    \hat{\psi}_2(r) = \frac{\sum_{a<b\in{\rm DD}}^{|r_{ab}| \in \mathrm{bin}} W_{\eta,\,a}\,W_{\eta,\,b}\,\eta_a\,\eta_b\,\cos\theta_{a}\,\cos\theta_b}{\sum_{a<b\in{\rm DD}}^{|r_{ab}| \in \mathrm{bin}} W_{\eta,\,a}\,W_{\eta,\,b}\,\cos\theta_a\,\cos\theta_b\,\cos\theta_{ab}} \, ,
\end{equation}
where we are performing a weighted sum over all unique pairs of galaxies following the notation of Equations \ref{eq:amodel} and \ref{eq:bmodel}. $\eta_{a,b}$ are the log-distance ratios measured for galaxy $a$ and $b$, and $W_{\eta,\, a}$ and $W_{\eta,\, b}$ are the weights associated with those galaxies. The angles $\cos\theta_a$, $\cos\theta_b$, $\cos\theta_{ab}$ define the geometry of the pair of galaxies with respect to our position: $\cos\theta_a = \hat{\vec{r}}\cdot\hat{\vec{r_a}}$, $\cos\theta_b = \hat{\vec{r}}\cdot\hat{\vec{r_b}}$, and $\cos\theta_{ab} = \hat{\vec{r_a}}\cdot\hat{\vec{r_b}}$. We refer the reader to Figure \ref{fig:geom} for a visual representation.

The estimators that we use in this analysis are formed from the weighted data-data (DD) pair counts, measured over all unique galaxy pairs in the data catalogue.  We apply a normalisation correction when generating the final momentum correlation function estimator (given that the inputs to the estimators are the log-distance values $\eta$). This is done by re-scaling the $\hat{\psi}_1(r)$ and $\hat{\psi}_2(r)$ measurements by $\alpha_{vv}(r)$, given by,
\begin{equation}
    \alpha_{vv}(r) = \frac{\sum_{a<b}^{|r_{ab}| \in \mathrm{bin}} W_{\eta,a} \,W_{\eta,b}}{\sum_{a<b}^{|r_{ab}| \in \mathrm{bin}} W_{\eta,a} \,W_{\eta,b}\,\,\alpha_a\,\alpha_b} \, ,
\end{equation}
where the weights are defined as above for a given pair of galaxies $a$ and $b$, and $\alpha_i$ corresponds to the conversion factor between velocity and log distance defined in Equation \ref{eq:alpha}.  We measure all of our correlations using bin widths of $6 \, h^{-1}$ Mpc, in the range $0 - 150 \, h^{-1}$ Mpc, with the python packages {\sc pycorr} and {\sc crosscorr} as described below.

\subsubsection{Galaxy auto-correlation estimator}

We use the Landy-Szalay estimator \citep{Landy1993} to measure the galaxy auto-correlation multipoles as a function of $r$, the separation between pairs of galaxies, and $\mu$, the cosine of the angle between the separation vector and the vector originating from the observer that bisects the angle $\cos\theta_{ab}$, $\cos\mu = \vec{r}.\vec{d}$:
\begin{equation}
    \hat{\xi}_{gg}(r,\mu) = \frac{DD(r,\mu) - 2DR(r,\mu) + RR(r,\mu)}{RR(r,\mu)} \, ,
\end{equation}
where $DD$ are the weighted `data-data' pair counts measured solely from unique galaxy pairs in the data catalogue containing $N_D$ galaxies, likewise $RR$ are the weighted `random-random' pair counts measured solely from unique pairs in the random catalogue containing $N_R$ galaxies, and $DR$ are the weighted `data-random' pair counts measured from the unique pairs drawn from the data and random catalogues. Each of these pair counts have been normalised to account for the differing sizes of the data and random catalogues such that,
\begin{equation}
    DD(r,\mu) = \frac{\sum_{a<b}^{(|r_{ab}|,\mu) \in \mathrm{bin}} W_{g,\,a}\,W_{g,\,b}}{(\sum_{N_D} W_g)^2} \, ,
\end{equation}
\begin{equation}
    DR(r,\mu) = \frac{\sum_{a<b}^{(|r_{ab}|,\mu) \in \mathrm{bin}} W_{g,\,a}\,W_{g,\,b}}{\sum_{N_D} W_g\cdot\sum_{N_R}W_g} \, ,
\end{equation}
\begin{equation}
    RR(r,\mu) = \frac{\sum_{a<b}^{(|r_{ab}|,\mu) \in \mathrm{bin}} W_{g,\,a}\,W_{g,\,b}}{(\sum_{N_R} W_g)^2} \, ,
\end{equation}
where $W_{g,\,a}$ and $W_{g,\,b}$ are the optimal galaxy density weights for galaxies $a$ and $b$.  The correlation function multipoles are calculated as,
\begin{equation}
    \hat{\xi}^{\ell}_{gg}(r) = \frac{2\ell + 1}{2}\int_0^1 d\mu \,\hat{\xi}_{gg}(r,\mu)\,L(\mu) \, ,
\label{eq:multipoles}
\end{equation}
with $\ell = 0, 2$ where we choose the width of each $\mu$ bin as $d\mu = 0.05$. We measure the two-point correlation functions using the python package {\sc pycorr}\footnote{\url{https://github.com/cosmodesi/pycorr}}, which is a wrapper for a modified version\footnote{\url{https://github.com/cosmodesi/Corrfunc}} of the python package {\sc corrfunc}\footnote{\url{https://github.com/manodeep/Corrfunc}} \citep{sinha2019, sinha2020}.  Finally, we measure and apply an integral constraint correction to the galaxy correlation functions, assuming the best-fitting correlation function model. The integral constraint introduces an additive bias, arising from the fact that we estimate the mean galaxy number density from the same survey used to measure $\xi_{gg}$. This causes the measured two-point correlation function to generally be biased low. We compute the required correction as
\begin{equation}
    IC = \frac{\sum RR(r)\,\xi_{gg}(r)}{\sum RR(r)}\,,
\end{equation}
where RR is as described and $\xi_{gg}$ is the best-fitting model for the two-point correlation function. The values of the integral constraint correction are $7.3 \times 10^{-4}$ and $6.2 \times 10^{-4}$ for the data and mock samples, respectively.

\subsubsection{Galaxy-momentum cross-correlation estimator}

The estimator for the galaxy-momentum cross-correlation function in bins of separation $r$ and angle to the line-of-sight $\mu$ is given by,
\begin{equation}
\begin{split}
    \hat{\xi}_{g\eta}(s,\mu) = \left(\frac{\sum_{N_R}W_{g\,}W_{\eta}}{\sum_{N_D}W_g\,W_{\eta}}\right)\frac{\sum_{a<b\in {\rm DD}} \,W_{g,a}\,W_{\eta,b}\,\eta_b}{\sum_{a<b\in {\rm RR}} W_{g,a}\,W_{\eta,b}} - \\
    \left(\frac{\sum_{N_R}W_{\eta}}{\sum_{N_D}W_{\eta}}\right)\frac{\sum_{a<b\in {\rm RD}} \,W_{g,a}\,W_{\eta,b} \,\eta_b}{\sum_{a<b\in {\rm RR}} W_{g,a}\,W_{\eta,b}} \, ,
\end{split}
\end{equation}
where we again perform a weighted sum over all unique pairs in the sample.  In the second term of the estimator, we subtract the momentum cross-correlation with the random galaxy catalogue, which acts to reduce the variance in our measurements by accounting for edge effects in the data \citep{Turner2021}.  This is analogous to how the Landy-Szalay estimator for the two-point correlation function improves over estimators such as those introduced by \cite{Peebles1974} or \cite{Davis1983}.  We convert the cross-correlation function measured in $(r,\mu)$ bins to the dipole $\xi_{g\eta}^1(r)$, using the equivalent of Equation \ref{eq:multipoles} with $\ell = 1$.

We apply an analogous normalisation to the cross-correlation as we do for the momentum auto-correlation function, in order to produce a measurement of the galaxy-momentum cross-correlation rather than the galaxy-log-distance ratio cross-correlation. This is performed by multiplying our measurement of $\hat{\xi}_{g\eta}^1$ by $\alpha_{gv}$, where
\begin{equation}
    \alpha_{gv}(r) = \frac{\sum_{a<b}^{|r_{ab}| \in \mathrm{bin}} W_{g,a} \,W_{\eta,b}}{\sum_{a<b}^{|r_{ab}| \in \mathrm{bin}} W_{g,a} \,W_{\eta,b}\,\,\alpha_b} \, ,
\end{equation}
where our galaxy and log-distance weights are defined as above, and we now only require one factor of $\alpha$ in the equation, as we are only converting one log-distance ratio to a velocity.

\subsubsection{Introducing {\sc crosscorr}}

We measure the momentum auto-correlations and galaxy-momentum cross-correlations using the python package {\sc crosscorr}\footnote{\url{https://github.com/r-jturner/crosscorr}}, created for this work. Similar to {\sc corrfunc}, {\sc crosscorr} provides a python wrapper for a set of C codes that compute correlation statistics. {\sc crosscorr} was created to compute the momentum correlations, because of their dependence on the previously described pair-specific geometry terms. {\sc crosscorr} is capable of measuring three-dimensional momentum correlation functions, useful for simulation cases where the full 3D momentum for each galaxy is known, and measuring the momentum correlations dependent on the radial component of galaxy velocities. The radial velocity correlations are the relevant computation for analyses of real data.  {\sc crosscorr} can measure correlations in both ($r$) and ($r$,$\mu$) space, and has the functionality to compute the multipoles of the correlation functions from the two-dimensional data.

{\sc crosscorr} is not as optimized for efficiency as {\sc corrfunc}, and the recommended usage is the scenario in which there are individual pairwise weights that must be computed, such as when determining the positions of pairs of galaxies relative to the observer and to themselves. To ensure that we were not introducing a systematic bias into our analysis with the addition of a separate codebase, we measured the two-point correlation function for an arbitrary selection of mocks using both {\sc corrfunc} and {\sc crosscorr}, and found our results to be identical. Based on this, we conclude that the use of both codes does not unduly influence our findings. We recommend the use of {\sc crosscorr} for computing the momentum correlation and cross-correlation statistics, and {\sc corrfunc} for computing galaxy clustering statistics.

\subsection{Minimisation}
\label{ssec:minim}

\begin{figure}
    \centering
    \includegraphics[width=0.9\linewidth]{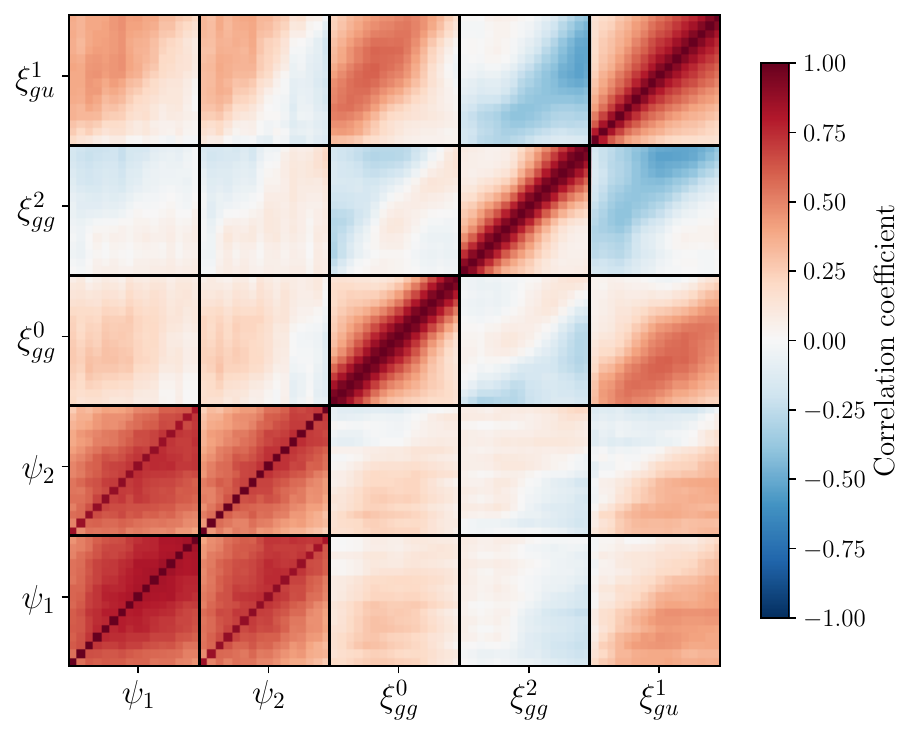}
    \caption{The correlation matrix used in this analysis, generated from all 675 mocks and using the fitting range $24 - 120 \, h^{-1}$ Mpc for each statistic. For a bin width of $6 \, h^{-1}$ Mpc this corresponds to 16 bins per statistic, and hence a matrix with dimensions $80 \times 80$. We employ a colour bar such that bins with higher correlation are redder, and bins that exhibit more anti-correlation are bluer. Bins on the diagonal are perfectly correlated by construction. Bins that exhibit no correlation or anti-correlation are white.}
    \label{fig:correlation}
\end{figure}

We now describe how we fit our model parameters to the measured correlation functions.  Following the approach of \cite{FeiPV}, we fit for five free parameters [$f\sigma_8$, $b_1\sigma_8$, $b_2\sigma_8$, $\sigma_{vT}^2$, $\sigma_{vS}^2$] using an MCMC framework.  We summarise these parameters as:
\begin{itemize}
    \item $f\sigma_8$: the normalised growth rate of large-scale structure,
    \item $b_1\sigma_8$: the normalised linear biasing parameter, 
    \item $b_2\sigma_8$: the normalised second-order local biasing parameter,
    \item $\sigma_{vT}^2$: the non-linear velocity dispersion for the loop terms $P_{02}$, $P_{04}$, $P_{12}$, $P_{22}$, and the vector part of $P_{13}$,
    \item $\sigma_{vS}^2$: the non-linear velocity dispersion for the loop term $P_{03}$, and the scalar part of $P_{13}$.
\end{itemize}
We assign uniform priors to each of these parameters, as listed in Table \ref{tab:priors}. 
\begin{table}[]
    \centering
    \begin{tabular}{c|c|c}
        Parameter & Unit & Prior \\
        \hline
        $f\sigma_8$      & -- & $\mathcal{U}(0,1.5)$\\
        $b_1\sigma_8$    & -- & $\mathcal{U}(0,3)$\\
        $b_2\sigma_8$    & -- & $\mathcal{U}(-5,5)$\\
        $\sigma_{vT}^2$ & $(h^{-1}$Mpc$)^2$ & $\mathcal{U}(0,350)$\\
        $\sigma_{vS}^2$ & $(h^{-1}$Mpc$)^2$ & $\mathcal{U}(0,350)$\\
    \end{tabular}
    \caption{A summary of the parameters considered in this analysis, and the uniform priors we place on each of them.}
    \label{tab:priors}
\end{table}
To perform our MCMC we use the python package {\sc{emcee}} \citep{emcee}, where the loss function that we minimise is the equation for chi-squared
\begin{equation}
    \chi^2(\theta) = (v_d - v_m(\theta))^T \,\hat{C}^{-1}\,(v_d - v_m(\theta)) \, ,
\end{equation}
where $v_d$ is the vector containing the data measurements (the concatenated set of correlation functions), $v_m$ is the vector containing the model predictions for the estimators for a given set of parameters $\theta$, and $\hat{C}^{-1}$ is the inverse of the covariance matrix constructed from the 675 mocks described in Section \ref{sec:mocks}. We account for the limited number of mocks used in creating the covariance matrix by applying the corrective Hartlap factor \citep{2007A&A...464..399H},
\begin{equation}
    \hat{C}^{-1} = \frac{n - p -2}{n - 1}\,C^{-1} \, ,
\end{equation}
where $C^{-1}$ is the original inverse covariance matrix, $n = 675$ is the number of mocks used in its construction, and $p$ is the length of our data vector, $v_d$. Applying this correction accounts for the additional noise introduced into our analysis when inverting the matrix constructed from limited mocks.  The corrected correlation matrix corresponding to the covariance is shown in Figure \ref{fig:correlation}.

We run MCMC analyses to obtain the posterior likelihoods for our parameters, placing 100 walkers in our parameter space and running a 5000 step analysis. 1500 steps are used for burn-in, and the other 3500 steps are used for analysis.  %We find that using longer chains did not affect the robustness of our results nor the convergence of the chains. 
We find that these chains converge well, finding that the $\hat{R}$ statistic \citep{Gelman1992} for $f\sigma_8$ is less than 1.03, and for all five parameters $\hat{R} < 1.10$.  
In cases where running MCMC analyses proves either too computationally expensive or time-consuming, we instead made use of the minimisation algorithms in {\sc scipy} to find the combination of parameters that minimise our $\chi^2$. When optimising in this manner, we set our initial guesses for each parameter to $[0.465, 0.811, 0.000, 100.0, 100.0]$, noting that our results do not depend on this choice.

\begin{figure*}
    \centering
    \includegraphics[width=0.9\linewidth]{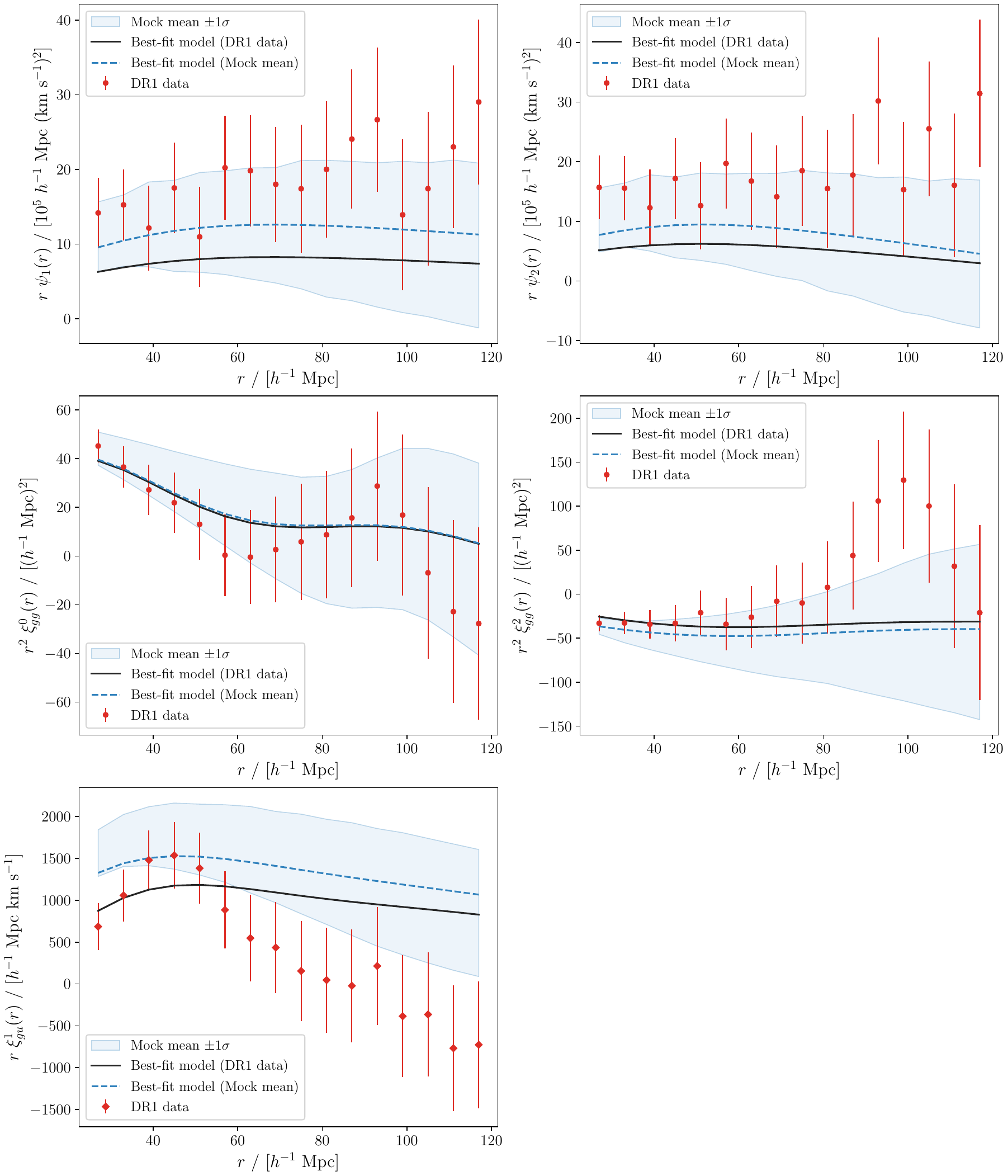}
    \caption{The five mock mean and data correlation functions that we consider in this analysis, and the best-fitting models for each.  The light blue shaded region depicts the mock mean and $1\sigma$ errors as determined from all 675 mocks. The red points represents the measurements made from the DESI DR1 data, using BGS clustering data, FP velocity data, and the $z$-cut TF velocity data. The errors on these points are the same $1\sigma$ errors determined from the mocks. The black lines represent the best-fitting models generated using the parameters derived from the data, as shown in Figure \ref{fig:data_corner}, and the dashed blue lines represent the best-fitting models generated using the parameters derived from the mock mean, as shown in Figure \ref{fig:mock_corner}. We remind readers of the difference in the redshifts of the two datasets when interpreting this figure.}
    \label{fig:model-v-data-v-mock}
\end{figure*}

\section{Results}
\label{sec:results}

\subsection{Correlation functions}

In Figure \ref{fig:model-v-data-v-mock}, we present our measurements of the five momentum and galaxy correlation functions of the DESI DR1 data (red errorbars) and mock mean (light blue shaded regions).  The best-fitting models for the data and mocks are shown in solid black and dashed blue lines, respectively.  The standard deviation across the ensemble of mock measurements is indicated as the errors on the data.  Visually, we observe offsets between the DR1 momentum correlation measurements and their corresponding model predictions. While there may be some contribution from systematics causing this offset, we attribute most of the effect to the high degree of correlation present in all five statistics. This is most prevalent for the momentum correlation functions which are highly correlated in different scale bins (see Figure~\ref{fig:correlation}), and are therefore relatively insensitive to overall shifts across scale. We cannot rule out contributions due to systematic effects, which we hope to address this with the more extensive and homogeneous DR2 sample, but we are confident that our best-fitting models provide a good fit to the data as assessed by the chi-squared values we report in Table \ref{tab:data_v_stats_table}. By contrast, we attribute the behaviour of the clustering monopole and quadrupole measurements entirely to the correlation between separation bins.

%As can be seen from Table~\ref{tab:data_v_stats_table}, the velocity data alone favour large values of $f\sigma_8$, which are then mitigated when clustering data are included. This indicates that the limited footprint of the DR1 data may also be driving the enhanced momentum correlations relative to model predictions. We will investigate this potential effect when analysing the DR2 data, which will be significantly more homogeneous than DR1. 

In addition, we observe an amplitude offset between the best-fitting models derived from the DR1 data and the mock mean, which is most pronounced in the models with greater sensitivity to the peculiar velocity data. We attribute this offset to the mismatch in redshift between the data and mock samples, since we expect a larger value of the growth rate at $z=0.20$ than at $z=0.07$. Given the uncertainties in our results, we do not expect the impact of this difference to be significant to our analysis.
%We expect a larger value of the growth rate at $z = 0.20$ than at $z = 0.07$, leading to the higher amplitude of the mock mean models.

In the following sections, we will systematically consider the minimum $\chi^2$ statistic of the best-fitting models (using the full covariance), and favoured parameter values, when applying the fitting methodology discussed in Section \ref{sec:fitting} to the mocks and data described in Section \ref{sec:DESI}.

\subsection{Fits to mocks}
\label{ssec:mockresults}

\begin{figure*}
    \centering
    \includegraphics[width=0.9\linewidth]{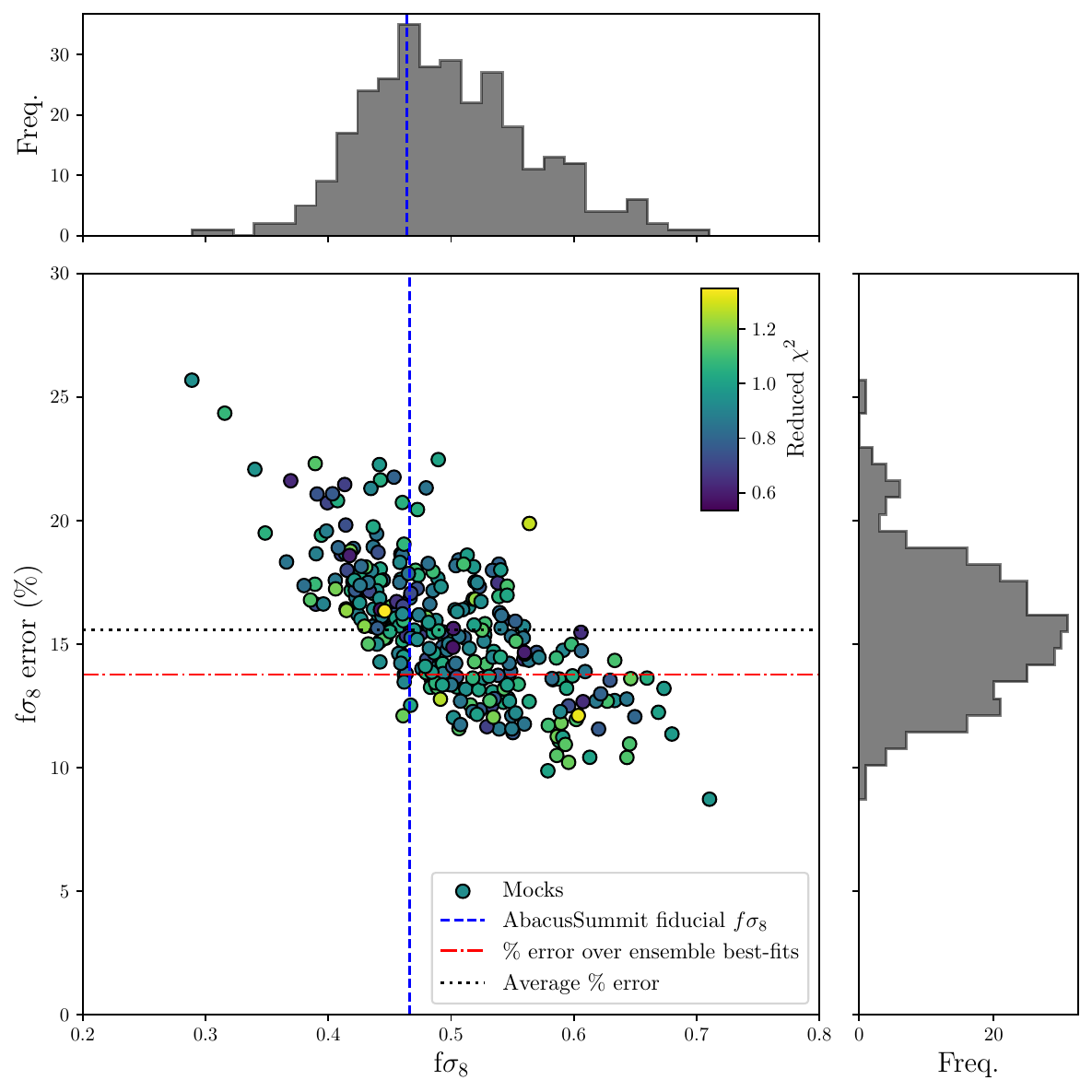}
    \caption{The best-fitting values of $f\sigma_8$ for 300 of the {\sc AbacusSummit} clustering and velocity mocks, plotted against the error in those values obtained from an MCMC analysis. The mock data includes both FP and TF velocity samples, and all five correlation function statistics were used to fit for the model parameters within a fitting range 24 - 120 $h^{-1}$ Mpc.  The points are colour-coded by the minimum reduced $\chi^2$ value, as indicated by the colour bar in the upper left of the plot. The vertical dashed line represents the fiducial value of $f\sigma_8$ in the {\sc AbacusSummit} mocks at the snapshot redshift $z = 0.20$. The dot-dashed line in red represents the ensemble error as a percentage, calculated as the standard deviation across the 300 best-fitting values of $f\sigma_8$. The dotted line in black represents the mean measurement error across the 300 mocks. The histograms on the top and right sides of the plot display the distribution of values of $f\sigma_8$ and $\sigma_{f\sigma_8}$, respectively.}
    \label{fig:675mock}
\end{figure*}

\begin{figure*}
    \centering
    \includegraphics[width=0.9\linewidth]{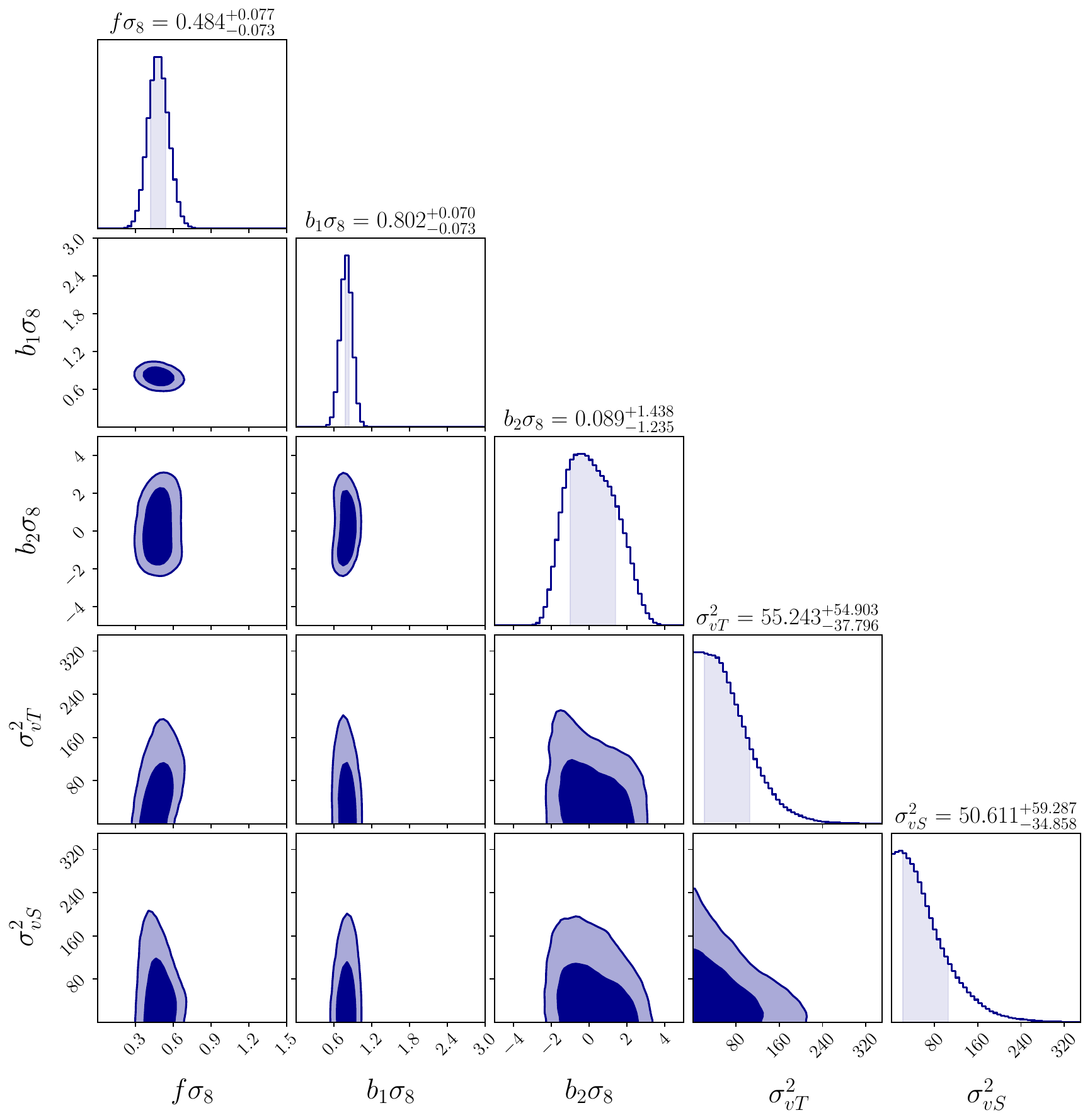}
    \caption{The joint confidence region for all parameters when fitting to the mock mean of the five correlation functions.  Differently shaded regions in the 2D contours show the $1\sigma$ and $2\sigma$ confidence intervals, and the three dashed lines in the 1D posterior probabilities indicate the 16th, 50th, and 84th quantiles, representing the $68\%$ confidence interval.}
    \label{fig:mock_corner}
\end{figure*}

\begin{figure*}
    \centering
    \includegraphics[width=0.9\linewidth]{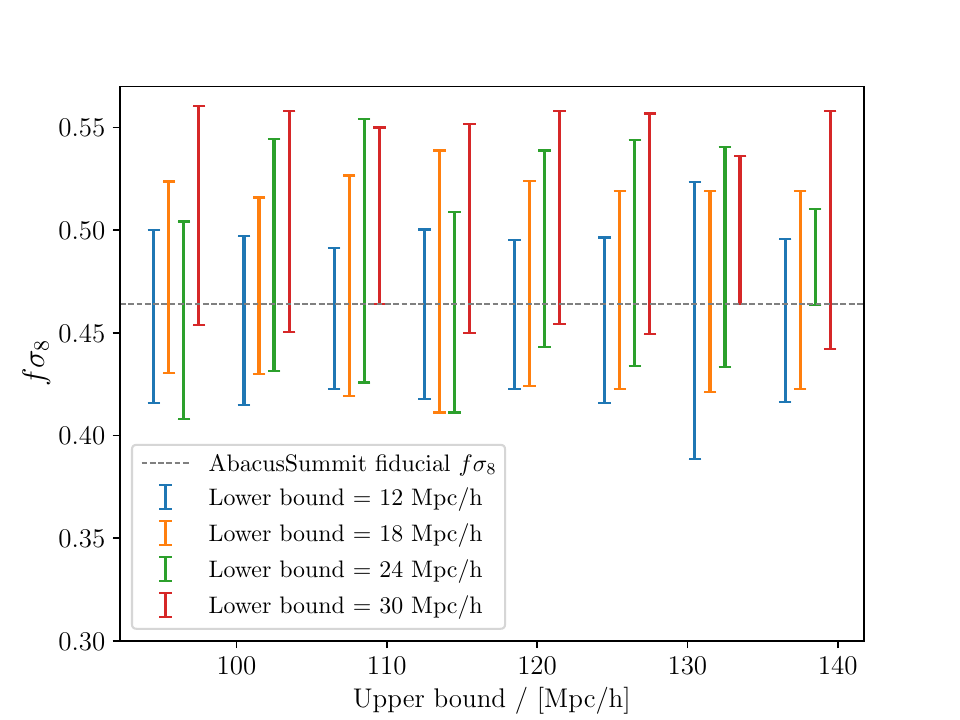}
    \caption{Measurements of $f\sigma_8$ obtained from a $\chi^2$ minimisation algorithm using the mock mean correlation functions, as a function of both the upper and lower bounds of the fitting range of separations. The upper bounds considered are [96, 102, 108, 112, 118, 124, 130, 136] $h^{-1}$ Mpc.  Lower bounds of 12 $h^{-1}$ Mpc are shown in blue, 18 $h^{-1}$ Mpc are shown in yellow, 24 $h^{-1}$ Mpc are shown in green, and 30 $h^{-1}$ Mpc are shown in red. The fiducial value of the growth rate in the {\sc AbacusSummit} simulations at $z = 0.2$ is shown by the horizontal dashed grey line.}
    \label{fig:mock_bound_error}
\end{figure*}

First, we wish to validate that our methodology recovers the fiducial cosmology of the mocks within acceptable accuracy.  We use the MCMC approach described above to determine the best-fitting set of parameters for 300 of the {\sc AbacusSummit} mocks.  We perform this test using all five correlation function statistics, for the fiducial combined velocity sample (the FP sample and $z$-cut TF sample), fitting for our five free parameters within the separation range 24 - 120 $h^{-1}$ Mpc.  (We will test below for the dependence of the results on the adopted fitting range.)

The results of this test are shown in Figure \ref{fig:675mock}, where we plot the best-fit value of $f\sigma_8$ for each mock against the estimated error in that measurement.  The distribution of best fit values of $f\sigma_8$, given by the histogram at the top of the figure, captures the fiducial value of the normalised growth rate $f\sigma_8 = 0.462$ in the {\sc AbacusSummit} mocks at redshift $z = 0.20$, as indicated by the vertical blue line, validating our methodology. Each point in Figure \ref{fig:675mock} is colour-coded based on the minimum reduced $\chi^2$, and we can see that our mock fits have an acceptable goodness-of-fit. The average error derived from this approach across the 300 mocks is approximately $15.6\%$, as given by the dot-dashed red line, while the standard deviation in the best-fitting $f\sigma_8$ values is closer to $13.8\%$ as given by the dotted black line. From these results we can assume that our methodology is unbiased and is capable of recovering the fiducial cosmology of the sample.
%The errors in these measurements are distributed around $10\%$. This is lower than the average error across the ensemble, which is approximately $16\%$, as indicated by the horizontal line. This is to be expected, as these forecast errors are under-estimates relative to a full MCMC fit, which we will adopt when analysing the real data.  

When we repeat the analysis using the ensemble mean of all correlation function measurements over the 675 mocks and the covariance for individual mocks, we find $f\sigma_8 = 0.484^{+0.077}_{-0.073}$, as shown in Figure \ref{fig:mock_corner}. This represents a measurement error of $15.5\%$, roughly in line with the average measurement error indicated in Figure \ref{fig:675mock}. The best-fitting parameters obtained here are used to produce the models shown in blue in Figure \ref{fig:model-v-data-v-mock}.

We also tested if our results have any significant dependence on the fitting range of separations we utilise.  Again, we fit to the mock mean results for all five correlation statistics.  The results of this test are shown in Figure \ref{fig:mock_bound_error}, where we display the mean and $1\sigma$ error of our recovered $f\sigma_8$ fits as the upper and lower bounds of our fitting range vary.  We consider [12, 18, 24, 30] $h^{-1}$ Mpc for the lower limit of our fit, and [96, 102, 108, 112, 118, 124, 130, 136] $h^{-1}$ Mpc as the upper bound, using the same minimisation method as in Figure \ref{fig:675mock}.  We find that all $f\sigma_8$ measurements agree within the errors and recover the fiducial mock value of the growth rate.

We see a consistent trend (albeit within the statistical error margin) of the $f\sigma_8$ value with the lower bound considered, which can be seen by comparing the run of blue, yellow, green, and red error bars in the plot, for a given choice of upper bound. Whilst using a lower bound of 12 $h^{-1}$ Mpc provides measurements that are strictly closest to the fiducial cosmology, we also consider the $\chi^2$ values for each combination of fitting ranges.  These results show that using a lower bound of 12 $h^{-1}$ Mpc leads to a poor goodness-of-fit.  Taking all this information into account, we adopt a fitting range $24 - 120 \, h^{-1}$ Mpc for the rest of our analysis, which is capable of successfully recovering the fiducial mock value of $f\sigma_8$ with a reasonable $\chi^2$ value.

%\begin{figure}
%    \centering
%    \includegraphics[width=0.95\linewidth]{plots/mocks_optim_boundstest_chisq.png}
%    \caption{The reduced $\chi^2$ values associated with each of the growth rate fits shown in Figure \ref{fig:mock_bound_error}, varying the upper and lower bounds of the fitting range.  Each cell is colour-coded according to the value of $\chi^2$, as given by the colour bar on the right-hand side of the figure, and the value of $\chi^2$ is also shown in the centre of the cell.}
%    \label{fig:mock_bound_chisq}
%\end{figure}

\subsection{Fits to data}
\label{ssec:dataresults}

\begin{figure*}
    \centering
    \includegraphics[width=0.9\linewidth]{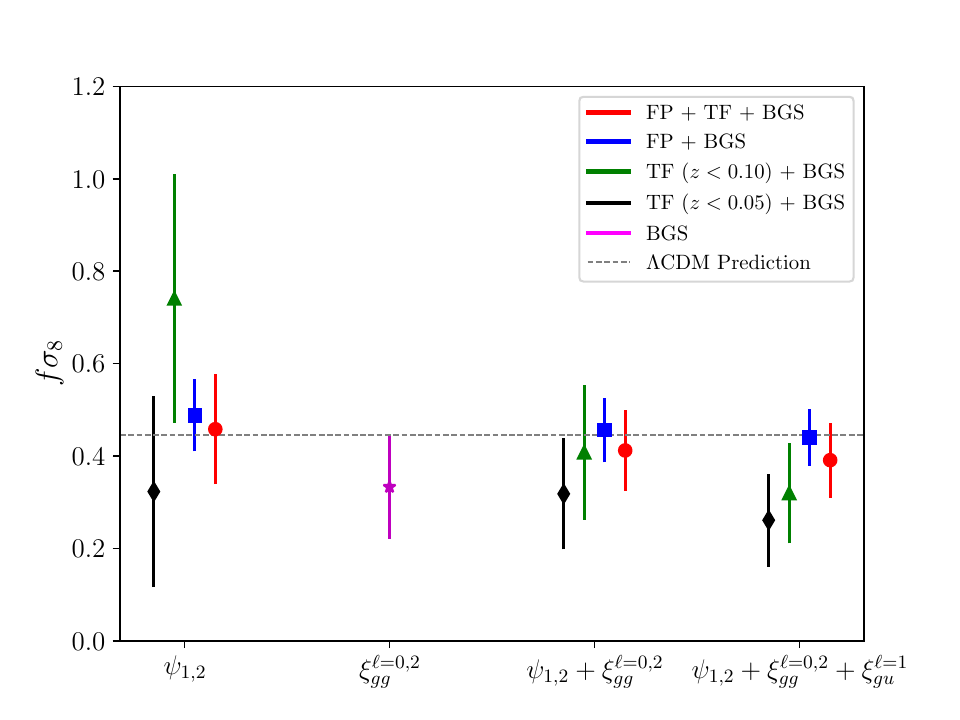}
    \caption{Measurements of $f\sigma_8$ using different combinations of input data, and fitting to different combinations of correlation statistics.  From left to right, we fit using only the momentum correlations [$\psi_1 + \psi_2$], only the galaxy clustering correlations [$\xi_{gg}^0 + \xi_{gg}^2$], the momentum and galaxy auto-correlations together [$\psi_1 + \psi_2$ + $\xi_{gg}^0 + \xi_{gg}^2$], and finally all five statistics [$\psi_1 + \psi_2$ + $\xi_{gg}^0 + \xi_{gg}^2 + \xi_{gu}^1$].  Results displayed as red circles correspond to our canonical analysis using the BGS, FP and TF $z$-cut datasets, and the other measurements correspond to different choices of the velocity samples in combination with BGS: the FP sample only (blue squares), the full TF sample (green triangles), the $z$-cut TF sample (black diamonds), and a clustering-only analysis (purple stars).  The horizontal dashed grey line represents the \textit{Planck}+$\Lambda$CDM prediction for $f\sigma_8$ at the mean redshift of the full DESI PV DR1 sample, $z_{\rm eff} = 0.07$. These results are also tabulated in Table \ref{tab:data_v_stats_table}.}
    \label{fig:datafit_v_stats}
\end{figure*}

\begin{figure*}
    \centering
    \includegraphics[width=0.9\linewidth]{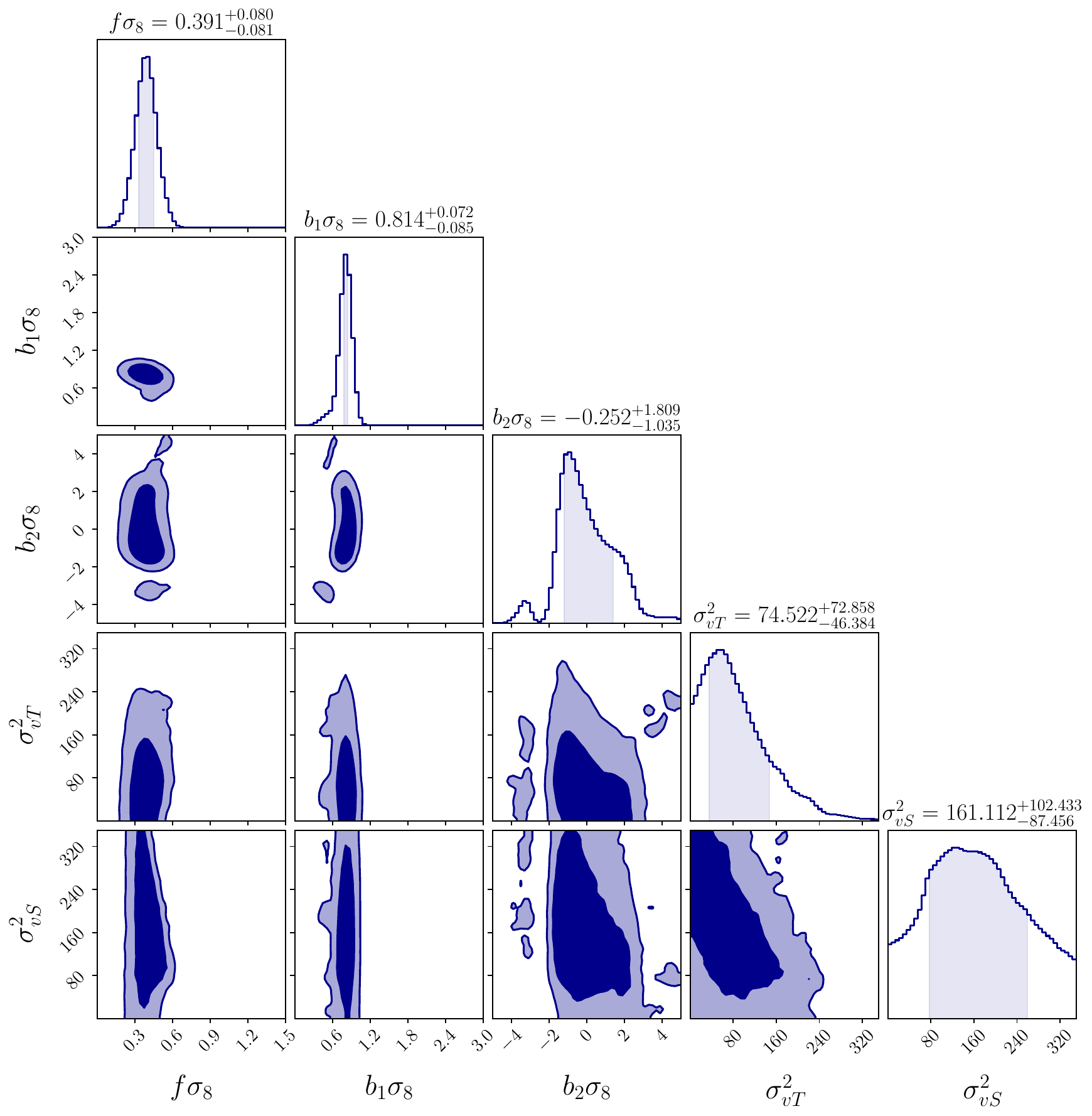}
    \caption{The joint confidence regions for all parameters when fitting to the five correlation functions of the BGS + FP + TF $z$-cut samples for the DESI DR1 dataset.  Differently shaded regions in the 2D contours show the $1\sigma$ and $2\sigma$ confidence intervals, and the three dashed lines in the 1D posterior probabilities indicate the 16th, 50th, and 84th quantiles, representing the $68\%$ confidence interval.}
    \label{fig:data_corner}
\end{figure*}

\begin{figure*}
    \centering
    \includegraphics[width=0.9\linewidth]{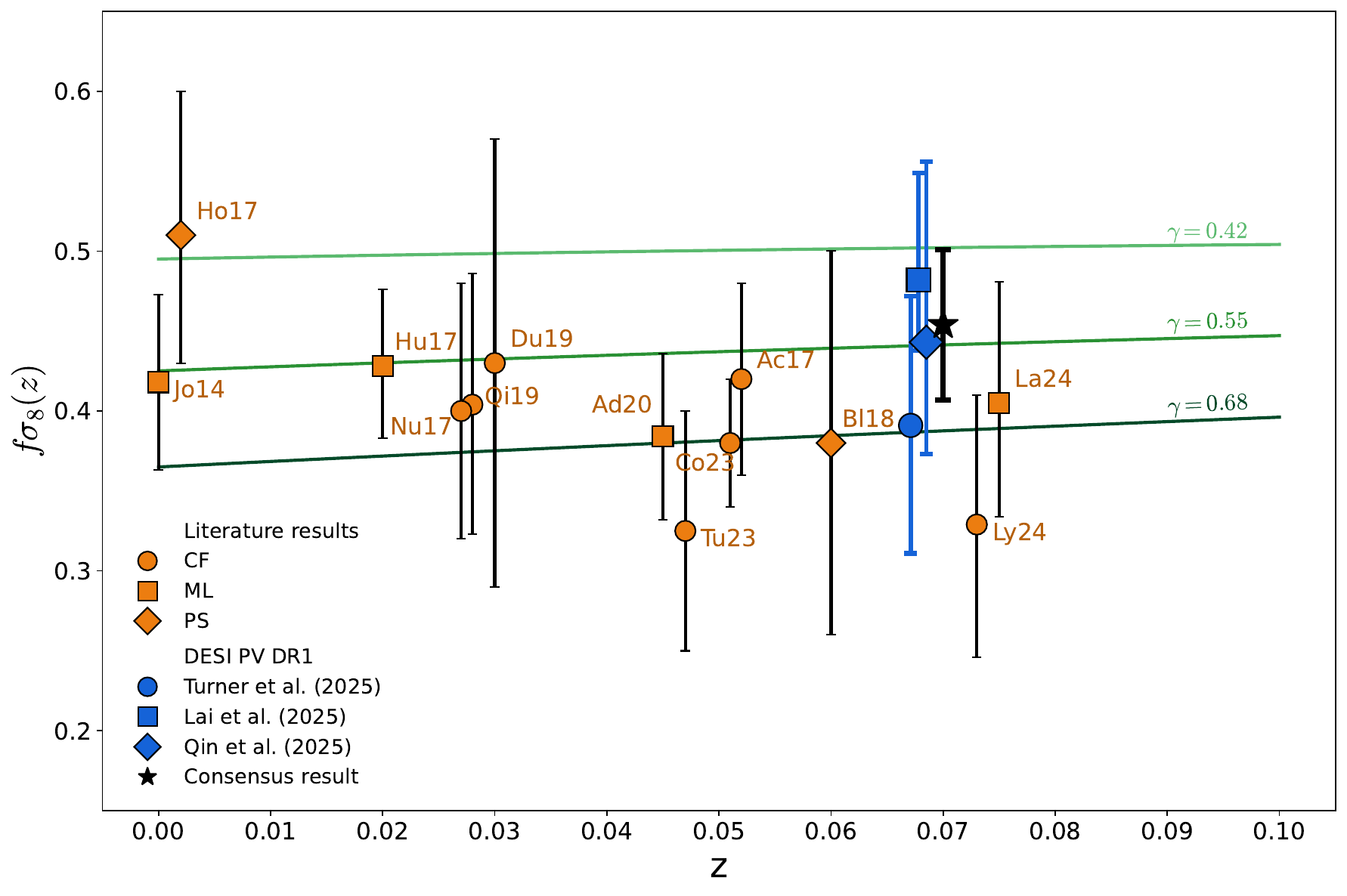}
    \caption{Various measurements of $f\sigma_8$, as a function of the effective redshifts of their respective datasets. Measurements may be slightly moved from their effective redshift for visual clarity. The three individual results from the DESI DR1 PV survey are shown in blue, and the combined consensus determination of $f\sigma_8$ from these results is shown in black and marked by a star. Other results from the literature obtained using the same or similar techniques are shown in orange. Results marked with circles are correlation function (CF) measurements, power spectrum (PS) measurements are shown as diamonds, and maximum-likelihood fields (ML) measurements are shown as squares. The three green lines represent three different predictions for $f\sigma_8(z)$, parameterised by $\gamma = [0.42,0.55,0.68]$, where $\gamma = 0.55$ is the value of the growth index for GR. The literature results shown from left to right are \cite{ Johnson2014, Howlett2017, Huterer2017, Nusser2017, Qin2019, Dupuy2019, Adams2020, Turner2023, Courtois2023, Achitouv2017, Blake2018, Lyall2024, Lai2023}. This information is also provided in Table \ref{tab:fs8-comparison}.}
    \label{fig:fs8_comp}
\end{figure*}

\begin{table*}[]
    \centering
    \renewcommand{\arraystretch}{1.7}
    \begin{tabular}{c|c|c|c}
        Data  & Statistics & $f\sigma_8$ & $\chi^2 / \nu$ \\
        \hline
        BGS + FP + TF ($z<0.05$) & $\psi_1 + \psi_2 + \xi_{gg}^0 + \xi_{gg}^2 + \xi_{gu}^1$ & $0.391^{+0.080}_{-0.081}$ & $91.140/75 = 1.215$ \\
        BGS + FP                 & $\psi_1 + \psi_2 + \xi_{gg}^0 + \xi_{gg}^2 + \xi_{gu}^1$ & $0.440^{+0.064}_{-0.061}$ & $113.840/75 = 1.518$  \\
        BGS + TF ($z<0.10$)      & $\psi_1 + \psi_2 + \xi_{gg}^0 + \xi_{gg}^2 + \xi_{gu}^1$ & $0.320^{+0.141}_{-0.109}$ & $61.845/75 = 0.825$ \\
        BGS + TF ($z<0.05$)      & $\psi_1 + \psi_2 + \xi_{gg}^0 + \xi_{gg}^2 + \xi_{gu}^1$ & $0.261^{+0.130}_{-0.101}$ & $112.971/75 = 1.506$  \\
        \hline
        BGS + FP + TF ($z<0.05$) & $\psi_1 + \psi_2 + \xi_{gg}^0 + \xi_{gg}^2$ & $0.412^{+0.083}_{-0.087}$ & $65.857/59 = 1.116$ \\
        BGS + FP                 & $\psi_1 + \psi_2 + \xi_{gg}^0 + \xi_{gg}^2$ & $0.456^{+0.068}_{-0.070}$ & $66.719/59 = 1.131$ \\
        BGS + TF ($z<0.10$)      & $\psi_1 + \psi_2 + \xi_{gg}^0 + \xi_{gg}^2$ & $0.408^{+0.158}_{-0.146}$ & $51.837/59 = 0.879$ \\
        BGS + TF ($z<0.05$)      & $\psi_1 + \psi_2 + \xi_{gg}^0 + \xi_{gg}^2$ & $0.318^{+0.140}_{-0.120}$ & $110.853/59 = 1.879$ \\
        \hline
        FP + TF ($z<0.05$) & $\psi_1 + \psi_2$  & $0.458^{+0.099}_{-0.119}$ & $29.465/27=1.091$ \\
        FP                 & $\psi_1 + \psi_2$  & $0.488^{+0.071}_{-0.078}$ & $27.965/27=1.036$ \\
        TF ($z<0.10$)      & $\psi_1 + \psi_2$  & $0.741^{+0.193}_{-0.269}$ & $12.130/27=0.449$\\
        TF ($z<0.05$)      & $\psi_1 + \psi_2$  & $0.323^{+0.238}_{-0.207}$ & $80.796/27=2.992$ \\
        \hline
        BGS                & $\xi_{gg}^0 + \xi_{gg}^2$   & $0.333^{+0.145}_{-0.112}$ & $32.712/27=1.212$\\
    \end{tabular}
    \caption{Growth rate measurements obtained from the MCMC analysis described in Section \ref{sec:fitting}, fitting different combinations of correlation statistics for different combinations of the DESI DR1 datasets (as described in Section \ref{sec:results}). For each combination of statistics we fit for all five free parameters discussed in Section \ref{sec:fitting}. In all cases, the fitting range used to fit the models to the DESI DR1 data measurements is 24 - 120 $h^{-1}$ Mpc, and the fit is performed assuming the mean redshift of the full DESI PV DR1 sample. These results are also shown in Figure \ref{fig:datafit_v_stats}.}
    \label{tab:data_v_stats_table}
\end{table*}

We now turn our attention to analysing the DESI DR1 dataset.  In Figure \ref{fig:datafit_v_stats} we show $f\sigma_8$ determinations using different combinations of correlation function statistics and input datasets, which will be described in more detail throughout this subsection.  We consider four combinations of statistics:
\begin{enumerate}
    \item Only the momentum auto-correlation statistics ($\psi_1 + \psi_2$);
    \item Only the galaxy clustering multipoles ($\xi_{gg}^0 + \xi_{gg}^2$);
    \item Both the momentum correlation statistics and the clustering multipoles ($\psi_1 + \psi_2 + \xi_{gg}^0 + \xi_{gg}^2$);
    \item All five statistics including the cross-correlation dipole ($\psi_1 + \psi_2 + \xi_{gg}^0 + \xi_{gg}^2 + \xi_{gu}^1$).
\end{enumerate}
For each of these combinations, we fit for $f\sigma_8$ and the other parameters following our MCMC procedure, using the following sets of peculiar velocity data (and the same clustering sample in all cases):
\begin{enumerate}
    \item The FP velocity sample (shown in blue in Figure \ref{fig:datafit_v_stats});
    \item The TF velocity sample (in green);
    \item The TF velocity sample with a cut $z < 0.05$ (in black), which we consider motivated by the results described below;
    \item Combining the FP and $z$-cut TF samples (in red);
    \item Excluding velocity data and only considering clustering data (in purple).
\end{enumerate}
We list the corresponding $f\sigma_8$ measurements in Table \ref{tab:data_v_stats_table}, and summarise the results in the following sub-sections.

\subsubsection{Momentum auto-correlation fits}

Fitting to the $\psi_1 + \psi_2$ measurements from the FP dataset results in a growth rate constraint $f\sigma_8 = 0.488^{+0.071}_{-0.078}$.
%\cab{Need to check this error, since it seems too good compared to other errors reported below, does it correspond to the MCMC?}. 
We characterise the goodness-of-fit using the reduced chi-squared, which we find to be $\chi^2_{\nu} = 1.036$ for this case.  The full TF dataset favours a high growth rate, albeit with a large error, $f\sigma_8 = 0.741^{+0.193}_{-0.269}$. Applying a $z < 0.05$ redshift cut to the TF data lowers the best-fitting growth rate, where we now find $f\sigma_8 = 0.323^{+0.238}_{-0.207}$.  Augmenting the FP dataset with the $z$-cut TF sample, we find $f\sigma_8 = 0.458^{+0.099}_{-0.119}$ with a reduced chi-squared value of $\chi^2_{\nu} = 1.091$. Hence, we conclude that the current DR1 TF sample does not add significant information to the growth rate fit, although given the excess correlation we measure for the TF dataset, we adopt a conservative approach of including the $z < 0.05$ cut for the TF sample when analysing the combined PV dataset.

\subsubsection{Galaxy auto-correlation multipole fits}

Next, we consider fitting our correlation model to only the clustering multipoles $\xi_{gg}^0 + \xi_{gg}^2$, using the BGS $z<0.10$ redshift sample.  These fits prefer a slightly lower value for the growth rate than the momentum-only fits, $f\sigma_8 = 0.333^{+0.145}_{-0.112}$, with $\chi^2_{\nu} = 1.212$.  The precision of the growth rate determination from the clustering statistics is hence similar to that from the momentum correlation statistics, and the recovered values of $f\sigma_8$ are also consistent, when considering the statistical errors.  Hence, we now consider fits to combinations of these statistics.

\subsubsection{Momentum and galaxy auto-correlation fits}

We now combine the momentum and clustering correlation statistics to improve the accuracy of our growth rate determination.  For our fiducial analysis case we find $f\sigma_8 = 0.412^{+0.083}_{-0.087}$ with a goodness-of-fit of $\chi^2_{\nu} = 1.116$.  Comparing this result to our fit to solely the momentum statistics, our goodness-of-fit values are similar ($\chi^2_{\nu} = 1.116$ vs. $1.091$) but the overall error in our measurement is improved by $22.0\%$. Compared to the clustering-only fits, the total error in the combined momentum-clustering fit is improved by $33.9\%$. It is clear from these results that combining peculiar velocity data and galaxy clustering data can achieve more precise growth rate constraints than using individual samples.

\subsubsection{Full-correlation fits}

We now present results fitting to all correlation statistics, including the dipole of the momentum-galaxy cross-correlation, using the BGS sample and the combined velocity datasets.  We obtain $f\sigma_8 = 0.391^{+0.080}_{-0.081}$, with $\chi^2_{\nu} = 1.215$.  Compared with the joint momentum-galaxy fit described in the previous subsection, we find that the inclusion of the dipole further improves our growth rate error by $5.3\%$. This corresponds to an improvement of $37.4\%$ when compared to the clustering-only fit.  We consider this measurement of 
\begin{equation}
    f\sigma_8(z=0.07) = 0.391^{+0.080}_{-0.081} \,\,\rm{(20.6\% \,\,error)}
\end{equation}
to be our fiducial correlation function constraint on the growth rate from the DESI DR1 dataset.

In Figure \ref{fig:data_corner} we present the joint confidence regions from our MCMC analysis for all five parameters, for this fiducial analysis.  This result is constructed similarly to the mock mean results in Figure \ref{fig:mock_corner}, and the best-fitting parameters obtained here are used to produce the models shown in black in Figure \ref{fig:model-v-data-v-mock}.  We find that the preferred values of the astrophysical parameters, the galaxy bias and velocity dispersions, are similar for the mocks and data, whilst noting that the velocity dispersion parameters are more consistent with zero in the mock mean analysis as compared to the fits to the data, where our fit for $\sigma_{vS}^2$ is largely unconstrained.

\subsubsection{Comparison with other DESI DR1 measurements}

In Figure \ref{fig:fs8_comp} and Table \ref{tab:fs8-comparison} we show our growth rate constraint alongside accompanying results we are presenting as part of the DESI PV DR1 analysis, as well as other recent results from the literature. The DESI PV results are shown in blue, whilst literature results are orange. Literature results are placed at the effective redshift reported by the authors, and some are slightly shifted away from their redshift for ease of visibility. We also show $f\sigma_8$ (GR$+\Lambda$CDM) predictions for three different values of $\gamma = [0.42,0.55,0.68]$ defined in Equation \ref{eq:gamma}. Correlation function results (`CF') are indicated by circle markers, power spectrum (`PS') results are indicated by diamonds, and maximum-likelihood fields (`ML') results are indicated by squares.

The growth rate determinations of the three methods we use for the DESI DR1 PV analysis are consistent with one another. \cite{YanPV}, who use a maximum-likelihood fields approach, find $f\sigma_8(z=0.07) = 0.482^{+0.049}_{-0.026}$(stat.)$+\,0.018$(sys.). \cite{FeiPV}, using the density and momentum power spectra, find $f\sigma_8(z=0.07) = 0.443^{+0.113}_{-0.070}$, a result in between the values favoured by the correlation function and maximum-likelihood approaches.

We combined the $f\sigma_8$ determinations of the three different methods as applied to the DR1 datasets to form a consensus value.  We accounted for the correlations between the analyses using the approach of \cite{2017MNRAS.464.1493S}, designed to combine the results of correlation function and power spectrum studies of the Baryon Oscillation Spectroscopic Survey,\footnote{We used the code \url{https://github.com/TyannDB/Gacomb} for this purpose.} using growth rate fits of the three methods as applied to a consistent set of {\sc AbacusSummit} mocks.  We hence find a consensus determination $f\sigma_8(z=0.07) = 0.4497 \pm 0.0548$ ($12.2\%$ error), see Section 9 of \cite{JulianPV} for additional information. This combined result is represented by a black errorbar and marked with a star in Figure \ref{fig:fs8_comp}.  We find that the improvement when combining the methods is appreciable, given that the different methods access different underlying scales with different weights.

The consensus growth rate determination is consistent with the \textit{Planck} $\Lambda$CDM prediction measured at the effective redshift of our sample, $f\sigma_8(z=0.07) = 0.446$ \citep{PlanckCollab2018}.  At the 1$\sigma$ level, our measurement also agrees with the DESI full-shape redshift-space distortion analysis of the BGS sample, which finds $f\sigma_8(z=0.295) = 0.38 \pm 0.09$ \citep{2025JCAP...09..008A}. We do note that the BGS full-shape measurement is made in the redshift range $0.1 < z < 0.4$, and so accesses clustering information that we do not but does not use peculiar velocity data.

Jointly fitting our consensus low-redshift growth rate result and the DESI DR1 full-shape clustering dataset, we measure gravitational growth index $\gamma_{\rm L} = 0.580^{+0.110}_{-0.110}$ \citep{FeiPV}, in accordance with $\gamma_{\rm L} \approx 0.55$ as predicted by general relativity.

\begin{table}[]
    \centering
    \renewcommand{\arraystretch}{1.7}
    \begin{tabular}{c|c|c|c}
        Author(s) & $f\sigma_8$ & Method & $z_{\mathrm{eff}}$\\
        \hline
        \cite{Johnson2014}  & 0.418 $\pm$ 0.055 & ML & 0.000 \\
        \cite{Howlett2MTF}  & 0.510$^{+0.090}_{-0.080}$ & PS & 0.000 \\
        \cite{Huterer2017}  & 0.428$^{+0.048}_{-0.045}$ & ML & 0.020 \\
        \cite{Nusser2017}   & 0.400 $\pm$ 0.080 & CF & 0.027 \\
        \cite{Qin2019}      & 0.404$^{+0.082}_{-0.081}$ & CF & 0.028 \\
        \cite{Dupuy2019}    & 0.430 $\pm$ 0.140 & CF & 0.028 \\
        \cite{Adams2020}    & 0.384 $\pm$ 0.052 & ML & 0.045 \\
        \cite{Turner2023}   & 0.325 $\pm$ 0.075 & CF & 0.045 \\
        \cite{Courtois2023} & 0.380 $\pm$ 0.040 & CF & 0.051 \\
        \cite{Achitouv2017} & 0.420 $\pm$ 0.060 & CF & 0.052 \\
        \cite{Blake2018}    & 0.380 $\pm$ 0.120 & PS & 0.060 \\
        \cite{Lyall2024}    & 0.329$^{+0.081}_{-0.083}$ & CF & 0.073 \\
        \cite{Lai2023}      & 0.405$^{+0.076}_{-0.071}$ & ML & 0.073 \\ 
        \hline
        DESI DR1 PS  & 0.443$^{+0.113}_{-0.070}$ & PS & 0.07 \\
        DESI DR1 ML  & 0.482$^{+0.067}_{-0.044}$ & ML & 0.07 \\ 
        DESI DR1 CF  & 0.391$^{+0.080}_{-0.081}$ & CF & 0.07 \\ 
        \hline
        DESI DR1 Consensus & 0.450 $\pm$ 0.055 & - & 0.07\\
    \end{tabular}
    \caption{The measurements of $f\sigma_8$ shown in Fig. \ref{fig:fs8_comp} ordered by the effective redshift of the growth rate measurements, as reported by the authors or as surmised from the datasets used. We also report the methodology used to obtain each measurement as `ML` (maximum-likelihood), `PS' (power spectrum), or `CF' (correlation function). Results from the DESI PV DR1 analyses are given together, with the final consensus result shown at the bottom. Any differences between the reported $z_{\mathrm{eff}}$ and the placement of the measurements in Fig. \ref{fig:fs8_comp} are solely for visual clarity.}
    \label{tab:fs8-comparison}
\end{table}
\subsubsection{Comparison with literature studies}

Comparing with the results of related studies of galaxy and momentum correlation functions in the literature: \cite{Turner2023} found $f\sigma_8(z=0.045) = 0.358 \pm 0.075$ ($21\%$ error) from analysis of the 6-degree Field Galaxy Survey Peculiar Velocity and redshift sample, and \cite{Lyall2024} determined $f\sigma_8(z=0.073) = 0.329^{+0.081}_{-0.083}$ ($25\%$ error) using the Sloan Digital Sky Survey PV dataset.  We find that the error in our determination of $f\sigma_8$ from the correlation function of the DESI DR1 sample ($20.6\%$) is only marginally improved over these previous studies, whilst representing the largest sample analysed to date.  To interpret this result, we note that the sky coverage of the DR1 sample remains highly incomplete relative to the full DESI footprint, as observations are still progressing.  Hence, the increased number of velocity tracers (and reduced measurement noise) is offset by increased sample variance.  We anticipate significant improvements in these metrics once the analysis of DESI Data Release 2 catalogue is complete.

\section{Conclusion}
\label{sec:conclusion}

In this paper we have presented our determination of the normalised growth rate of large-scale structure from the two-point correlations between density tracers and velocity tracers, using the DESI DR1 Bright Galaxy Survey galaxy catalogue, and Fundamental Plane and Tully-Fisher peculiar velocity catalogues. Consisting of over $415{,}000$ galaxy redshifts, and over $76{,}000$ galaxy peculiar velocities, this is the largest sample of its kind ever analysed. We consider five correlation function statistics in our analysis and construct non-linear models to describe them, based on power spectrum multipoles generated using 1-loop Eulerian perturbation theory.  We employ mock datasets to validate the robustness of our analysis, and to show that our methodology is capable of producing unbiased results. We also use the mocks to explore the dependence of our analysis on the range of correlation function separations we use to fit our models to our measurements, demonstrating that our results are insensitive to the adopted fitting range, 

The final constraint on the growth rate from our correlation function study is $f\sigma_8 = 0.391^{+0.080}_{-0.081}$, which represents a $37.4\%$ improvement in the error over the equivalent result when only fitting to the clustering monopole and quadrupole, and a $5.3\%$ improvement over the equivalent result when fitting to the clustering multipoles as well as the momentum correlation statistics.  Whilst our result does not represent a significant improvement in the growth rate error compared to previous samples, we attribute this to the incomplete nature of the DESI DR1 dataset, which impacts the overall effective volume of our analysis. Future analyses of the corresponding galaxy and momentum catalogues in DESI DR2, which covers a wider sky area, will further improve these results.   

The consensus determination from the DR1 samples, combining results from the correlation function, power spectrum, and maximum-likelihood fields methods, is $f\sigma_8(z = 0.07) = 0.4497 \pm 0.0548$.  The accuracy of our growth rate fits already exceeds equivalent results reported for the full DR1 BGS redshift-space distortion studies, demonstrating the potential of peculiar velocity measurements to contribute to testing gravitational physics in the local universe, and hence yielding important information on the properties of dark energy.

\section*{Data availability}

The DESI PV DR1 data products will be made available, upon acceptance of the papers in this series, at \url{https://data.desi.lbl.gov/doc/releases/dr1/}. Data points for all relevant figures in this paper are available at \url{https://doi.org/10.5281/zenodo.17668481}.

\section*{Acknowledgement}

RJT and CB acknowledge financial support received through Australian Research Council Discovery Project DP220101610.

This research made use of {\sc Astropy} \citep{astropy2013,astropy2018}, {\sc NumPy} \citep{numpy1, numpy2}, and {\sc SciPy} \citep{scipy}. Plots in this paper were produced using {\sc matplotlib} \citep{matplotlib} and {\sc corner} \citep{corner}.

This material is based upon work supported by the U.S. Department of Energy (DOE), Office of Science, Office of High-Energy Physics, under Contract No. DE–AC02–05CH11231, and by the National Energy Research Scientific Computing Center, a DOE Office of Science User Facility under the same contract. Additional support for DESI was provided by the U.S. National Science Foundation (NSF), Division of Astronomical Sciences under Contract No. AST-0950945 to the NSF’s National Optical-Infrared Astronomy Research Laboratory; the Science and Technology Facilities Council of the United Kingdom; the Gordon and Betty Moore Foundation; the Heising-Simons Foundation; the French Alternative Energies and Atomic Energy Commission (CEA); the National Council of Humanities, Science and Technology of Mexico (CONAHCYT); the Ministry of Science, Innovation and Universities of Spain (MICIU/AEI/10.13039/501100011033), and by the DESI Member Institutions: \url{https://www.desi.lbl.gov/collaborating-institutions}. Any opinions, findings, and conclusions or recommendations expressed in this material are those of the author(s) and do not necessarily reflect the views of the U. S. National Science Foundation, the U. S. Department of Energy, or any of the listed funding agencies.

The authors are honored to be permitted to conduct scientific research on I'oligam Du'ag (Kitt Peak), a mountain with particular significance to the Tohono O’odham Nation.
%\end{acknowledgement}

\bibliography{reference}
\section*{Affiliations}
\scriptsize
\noindent
$^{1}$ Centre for Astrophysics \& Supercomputing, Swinburne University of Technology, P.O. Box 218, Hawthorn, VIC 3122, Australia\\
$^{2}$ Aix Marseille Univ, CNRS/IN2P3, CPPM, Marseille, France\\
$^{3}$ Lawrence Berkeley National Laboratory, 1 Cyclotron Road, Berkeley, CA 94720, USA\\
$^{4}$ Department of Physics, Boston University, 590 Commonwealth Avenue, Boston, MA 02215 USA\\
$^{5}$ Department of Physics, Carnegie Mellon University, 5000 Forbes Avenue, Pittsburgh, PA 15213, USA\\
$^{6}$ Department of Physics \& Astronomy, University of Rochester, 206 Bausch and Lomb Hall, P.O. Box 270171, Rochester, NY 14627-0171, USA\\
$^{7}$ Dipartimento di Fisica ``Aldo Pontremoli'', Universit\`a degli Studi di Milano, Via Celoria 16, I-20133 Milano, Italy\\
$^{8}$ INAF-Osservatorio Astronomico di Brera, Via Brera 28, 20122 Milano, Italy\\
$^{9}$ Department of Physics \& Astronomy, University College London, Gower Street, London, WC1E 6BT, UK\\
$^{10}$ Korea Astronomy and Space Science Institute, 776, Daedeokdae-ro, Yuseong-gu, Daejeon 34055, Republic of Korea\\
$^{11}$ Instituto de F\'{\i}sica, Universidad Nacional Aut\'{o}noma de M\'{e}xico,  Circuito de la Investigaci\'{o}n Cient\'{\i}fica, Ciudad Universitaria, Cd. de M\'{e}xico  C.~P.~04510,  M\'{e}xico\\
$^{12}$ University of California, Berkeley, 110 Sproul Hall \#5800 Berkeley, CA 94720, USA\\
$^{13}$ Institut de F\'{i}sica d’Altes Energies (IFAE), The Barcelona Institute of Science and Technology, Edifici Cn, Campus UAB, 08193, Bellaterra (Barcelona), Spain\\
$^{14}$ Departamento de F\'isica, Universidad de los Andes, Cra. 1 No. 18A-10, Edificio Ip, CP 111711, Bogot\'a, Colombia\\
$^{15}$ Observatorio Astron\'omico, Universidad de los Andes, Cra. 1 No. 18A-10, Edificio H, CP 111711 Bogot\'a, Colombia\\
$^{16}$ Institut d'Estudis Espacials de Catalunya (IEEC), c/ Esteve Terradas 1, Edifici RDIT, Campus PMT-UPC, 08860 Castelldefels, Spain\\
$^{17}$ Institute of Cosmology and Gravitation, University of Portsmouth, Dennis Sciama Building, Portsmouth, PO1 3FX, UK\\
$^{18}$ Institute of Space Sciences, ICE-CSIC, Campus UAB, Carrer de Can Magrans s/n, 08913 Bellaterra, Barcelona, Spain\\
$^{19}$ University of Virginia, Department of Astronomy, Charlottesville, VA 22904, USA\\
$^{20}$ Fermi National Accelerator Laboratory, PO Box 500, Batavia, IL 60510, USA\\
$^{21}$ Institut d'Astrophysique de Paris. 98 bis boulevard Arago. 75014 Paris, France\\
$^{22}$ IRFU, CEA, Universit\'{e} Paris-Saclay, F-91191 Gif-sur-Yvette, France\\
$^{23}$ Center for Cosmology and AstroParticle Physics, The Ohio State University, 191 West Woodruff Avenue, Columbus, OH 43210, USA\\
$^{24}$ Department of Physics, The Ohio State University, 191 West Woodruff Avenue, Columbus, OH 43210, USA\\
$^{25}$ The Ohio State University, Columbus, 43210 OH, USA\\
$^{26}$ School of Mathematics and Physics, University of Queensland, Brisbane, QLD 4072, Australia\\
$^{27}$ Department of Physics, University of Michigan, 450 Church Street, Ann Arbor, MI 48109, USA\\
$^{28}$ University of Michigan, 500 S. State Street, Ann Arbor, MI 48109, USA\\
$^{29}$ Department of Physics, The University of Texas at Dallas, 800 W. Campbell Rd., Richardson, TX 75080, USA\\
$^{30}$ NSF NOIRLab, 950 N. Cherry Ave., Tucson, AZ 85719, USA\\
$^{31}$ Department of Physics, Southern Methodist University, 3215 Daniel Avenue, Dallas, TX 75275, USA\\
$^{32}$ Department of Physics and Astronomy, University of California, Irvine, 92697, USA\\
$^{33}$ Steward Observatory, University of Arizona, 933 N. Cherry Avenue, Tucson, AZ 85721, USA\\
$^{34}$ Sorbonne Universit\'{e}, CNRS/IN2P3, Laboratoire de Physique Nucl\'{e}aire et de Hautes Energies (LPNHE), FR-75005 Paris, France\\
$^{35}$ Department of Astronomy and Astrophysics, UCO/Lick Observatory, University of California, 1156 High Street, Santa Cruz, CA 95064, USA\\
$^{36}$ Department of Astronomy and Astrophysics, University of California, Santa Cruz, 1156 High Street, Santa Cruz, CA 95065, USA\\
$^{37}$ Departament de F\'{i}sica, Serra H\'{u}nter, Universitat Aut\`{o}noma de Barcelona, 08193 Bellaterra (Barcelona), Spain\\
$^{38}$ Instituci\'{o} Catalana de Recerca i Estudis Avan\c{c}ats, Passeig de Llu\'{\i}s Companys, 23, 08010 Barcelona, Spain\\
$^{39}$ Department of Physics and Astronomy, Siena University, 515 Loudon Road, Loudonville, NY 12211, USA\\
$^{40}$ Department of Physics and Astronomy, University of Waterloo, 200 University Ave W, Waterloo, ON N2L 3G1, Canada\\
$^{41}$ Perimeter Institute for Theoretical Physics, 31 Caroline St. North, Waterloo, ON N2L 2Y5, Canada\\
$^{42}$ Waterloo Centre for Astrophysics, University of Waterloo, 200 University Ave W, Waterloo, ON N2L 3G1, Canada\\
$^{43}$ Space Sciences Laboratory, University of California, Berkeley, 7 Gauss Way, Berkeley, CA  94720, USA\\
$^{44}$ Instituto de Astrof\'{i}sica de Andaluc\'{i}a (CSIC), Glorieta de la Astronom\'{i}a, s/n, E-18008 Granada, Spain\\
$^{45}$ Departament de F\'isica, EEBE, Universitat Polit\`ecnica de Catalunya, c/Eduard Maristany 10, 08930 Barcelona, Spain\\
$^{46}$ Department of Physics and Astronomy, Sejong University, 209 Neungdong-ro, Gwangjin-gu, Seoul 05006, Republic of Korea\\
$^{47}$ CIEMAT, Avenida Complutense 40, E-28040 Madrid, Spain\\
$^{48}$ National Astronomical Observatories, Chinese Academy of Sciences, A20 Datun Road, Chaoyang District, Beijing, 100101, P.~R.~China\\
\normalsize
%\appendix

\end{document}